\documentclass[traditabstract]{aa}

\usepackage{natbib}
\usepackage{graphicx}
\usepackage{amssymb}
\usepackage{amsmath}
\usepackage{color} 

\usepackage{txfonts}

\newcommand{\be}{\begin{equation}}
\newcommand{\ee}{\end{equation}}

\def\ba{\begin{eqnarray}}
\def\ea{\end{eqnarray}}

\def\msol{M_\odot}

\def\he3{^3He}

\def\ltsima{$\; \buildrel < \over \sim \;$}
\def\simlt{\lower.5ex\hbox{\ltsima}}
\def\gtsima{$\; \buildrel > \over \sim \;$}
\def\simgt{\lower.5ex\hbox{\gtsima}}
\def\vel{{\rm v}}
\def\velvec{{\rm \mathbf v}}
\def\dt{\delta T}

\newcommand{\cp}{\citep}
\newcommand{\ct}{\citet}

 \renewcommand{\vec}[1]{\mathbf{#1}}

\begin{document}

\title{Two-dimensional simulations of solar-like models with artificially enhanced luminosity -
I. Impact on convective penetration}
\titlerunning{2D simulations of solar-like models with artificially enhanced luminosity}

\author{I. Baraffe \inst{1,2}, J. Pratt\inst{3}, D. G. Vlaykov\inst{1}, T. Guillet\inst{1},T. Goffrey\inst{4},  A. Le Saux \inst{1}, T. Constantino \inst{1}}
\authorrunning{Baraffe et al.}

\offprints{I. Baraffe} 
   
\institute{
University of Exeter, Physics and Astronomy, EX4 4QL Exeter, UK
(\email{i.baraffe@ex.ac.uk})
\and
\'Ecole Normale Sup\'erieure, Lyon, CRAL (UMR CNRS 5574), Universit\'e de Lyon, France
\and
Department of Physics and Astronomy, Georgia State University, Atlanta GA 30303, USA
\and
Centre for Fusion, Space and Astrophysics, Department of Physics, University of Warwick, Coventry, CV4 7AL, UK
}

\date{}

\abstract{We performed two-dimensional, fully compressible, time-implicit simulations of convection  in a solar-like model with the MUSIC code. Our main motivation is to explore the impact of a common tactic adopted in numerical simulations of convection that use realistic stellar conditions. This tactic is to artificially increase the luminosity and to modify the thermal diffusivity of the reference stellar model. This work  focuses on the impact of these modifications on convective penetration (or overshooting) at the base of the convective envelope of a solar-like model. We explore a range of enhancement factors for the energy input (or stellar luminosity) and confirm the increase in the characteristic overshooting depth with the increase in the energy input, as suggested by analytical models and by previous numerical simulations. 
We performed high-order moments analysis of the temperature fluctuations for moderate enhancement factors and find similar flow structure in the convective envelope and the penetration region, independently of the enhancement factor.
As a major finding, our results highlight the importance of the impact of penetrative downflows on the thermal background below the convective boundary. This is a result of compression and shear which induce local heating and thermal mixing. The artificial increase in the energy flux  intensifies the heating process by increasing the velocities in the convective zone and at the convective boundary, revealing a subtle connection between the local heating of the thermal background and the plume dynamics. This heating also increases  the efficiency of heat transport by radiation which may counterbalance further heating and helps to establish a steady state. We suggest that the modification of the thermal background by penetrative plumes  impacts
the width of the overshooting layer.
Additionally, our results suggest that an artificial modification of the radiative diffusivity in the overshooting layer, rather than only accelerating the thermal relaxation, could also alter the dynamics of the penetrating plumes and thus the width of the overshooting layer. Results from simulations with an artificial modification of the energy flux and of the thermal diffusivity should thus be regarded with caution if used to determine an overshooting distance. 
}


\maketitle

\section{Introduction}

One of the major uncertainties in stellar physics is the treatment of mixing taking place at convective boundaries. 
Convective motions do not  abruptly stop at the classical Schwarzschild boundary, but extend beyond it. 
The complex dynamics resulting from convective penetration in stable layers is a major process in stellar interiors that drives the transport of chemical species and heat, strongly affecting the structure and the evolution of many types of stars. This process is usually called  overshooting, convective penetration or  convective boundary mixing. 
The same complex dynamics can also drive the transport  of angular momentum, affecting the rotational evolution of stars, the generation of a magnetic field in their interior and their magnetic activity. 
Overshooting is one of the oldest unsolved problems of stellar structure and evolution theory \cp[e.g.][]{shaviv73} and   affects all stars that develop a convective envelope or core, meaning  all stars with a mass above  $\sim 0.4 \msol$. 

 Analytical and semianalytical approaches have been developed to describe this process and estimate the width of the overshooting layer \cp[e.g.][]{schmitt84, zahn91, rempel04}.
With the improvement of computational methods and resources, an increasing number of studies have been devoted to numerical simulations of convection and overshooting using realistic stellar conditions (geometry, luminosity, thermal diffusivity, equation of state, opacities, etc.).
A commonly used tactic to increase the efficiency and improve the stability of these simulations is to artificially increase the luminosity (or nuclear energy for convective cores or burning shells) and to modify the thermal diffusivity of the reference stellar model. This tactic is common and has been used, for example, in \ct[][]{rogers06, meakin07, tian09, brun11, rogers13, brun17, hotta17, cristini17, Edelmann19, horst20}. 
This approach is used to increase the Mach number of the convective flow, reducing the disparity between advective and acoustic timescales, improving the efficiency of time-explicit codes limited by the Courant-Friedrich-Levy constraint. It is also used to provide numerical stability or to accelerate the thermal relaxation.
But  no examination of its potentially far-reaching impact has  been conducted. \ct{rempel04} pointed out that enhanced energy flux in numerical simulations could lead to unrealistically vigourous convection, which could impact the properties of the overshooting layer and could explain some of the discrepancies between analytical models and numerical simulations. Numerical simulations 
also suggest an increase in the overshooting depth with increasing flux \cp{hotta17, kapyla19}. Determining scaling laws of the overshooting depth as a function of the energy input could thus allow for an extrapolation of the results to more realistic stellar conditions and help to estimate the overshooting depth in real stars, as suggested by  \ct[][]{hotta17}, for example, for the Sun.
But \ct{kapyla19} also shows that an artificial modification of the heat conductivity in the radiative and overshooting regions could  impact the overshooting process. In some computational studies, both the luminosity and the thermal diffusivity are enhanced by the same factor to ensure that the thermal structure is unchanged compared to the reference stellar structure and with the expectation that the larger thermal diffusivity counterbalances the larger energy flux. This procedure has been proposed as a way to provide a good representation of the true dynamics of the system \cp[e.g.][]{rogers06, tian09, rogers13}. But this  expectation has never been demonstrated.  Another expectation concerns internal waves, excited by convective motions and by flows penetrating the convective boundary. Simulations with artificially modified luminosity and thermal diffusivity are also used to perform the analysis of internal waves, either for convective envelopes \cp[e.g][]{rogers06,brun11, alvan14} or for convective cores \cp[e.g.][]{rogers17,Edelmann19,horst20}. 
None of these works have examined whether the wave spectrum  of a realistic star is accurately predicted by such simulations.

The modification of the energy flux and thermal diffusivity commonly performed in stellar hydrodynamics simulations of convection and overshooting thus raises several questions. Firstly, to which extent can scaling laws of the overshooting depth with luminosity be extrapolated down to realistic stellar luminosities? Secondly, does the enhancement of the luminosity and thermal diffusivity only impact the timescale of the simulation, that is accelerates it, or does it have other effects on the dynamics of the region of interest? Thirdly, do  such boosted models predict a realistic spectrum of internal waves generated at the convective boundary and propagating  in the stable region?

We attempt to address these questions by performing a suite of two-dimensional simulations for a solar-like star and  analysing the impacts of artificially enhanced luminosity and thermal diffusivity.  These experiments are restricted for now to two-dimension models,  which 
 allow for longer simulations in order to calculate accurate statistics, and the examination of multiple simulations to explore a wider parameter range.
We will discuss in \S \ref{conclusion} the impact of the dimensionality on our main results. 
In this work we examine the properties of convection and  convective penetration at the base of the convective envelope. Two follow-up studies are underway. The first one is devoted to the properties of internal waves generated at the convective boundary  (Le Saux et al., in prep). The second one is devoted to the analysis of the overshooting depth based on a combination of the simulations presented in this work and Lagrangian tracer particles (Guillet et al., in prep).

\section{Numerical simulations}
In this work we use the fully compressible time-implicit code MUSIC. A full description of MUSIC and of the time-implicit integration can be found in \ct{viallet11, viallet16, goffrey17}.  Here, we provide a brief description of its main characteristics. MUSIC
solves the inviscid Euler equations in the presence of external gravity and
thermal diffusion:

\begin{eqnarray}
\frac{\partial \rho}{\partial t} &=& - \vec \nabla \cdot (\rho \vec \velvec),\\
\frac{\partial \rho \vec \velvec}{\partial t} &=& - \vec \nabla \cdot (\rho \vec \velvec \otimes \vec \velvec)-\vec \nabla p + \rho \vec g,\\
\frac{\partial \rho e}{\partial t} &=& -\vec \nabla \cdot (\rho e \vec \velvec) - p \vec{ \nabla} \cdot \vec \velvec + \vec \nabla \cdot (\chi \vec \nabla T),
\end{eqnarray}
\noindent where $\rho$ is the density, $e$ the specific internal energy, $\vec
\vel$ the velocity, $p$ the gas pressure, $T$ the temperature, $\vec g$ the
gravitational acceleration, and $\chi$ the thermal conductivity. 
The symbol $\otimes$ is the outer product. All  hydrodynamical simulations presented in this work are performed assuming spherically symmetric gravitational acceleration that does not evolve with time, meaning that $\mathbf{g}$ is calculated at time $t$=0 using the initial density profile and remains constant with time. 
For the stellar model that we consider in this work, radiative transfer is the major heat transport that contributes to the thermal conductivity, which is given for photons by
\begin{equation}
\label{eq:chirad}
\chi = \frac{16 \sigma T^3}{3\kappa \rho},
\end{equation}

\noindent where $\kappa$ is the Rosseland mean opacity, and $\sigma$ the Stefan-Boltzmann constant. In the following,  thermal diffusivity and radiative diffusivity are used interchangeably.
Realistic stellar opacities and equation of states appropriate for the description of stellar interiors are implemented in MUSIC. Opacities are interpolated from the OPAL tables \cp{Iglesias96} for solar metallicity and the equation of state is based on the OPAL EOS tables of \ct{rogers02}, which are appropriate for the description of solar-like interior structures.  

\subsection{Initial stellar models}

\begin{figure}[h!]
\vspace{-0.5cm}
\includegraphics[height=8cm,width=8cm]{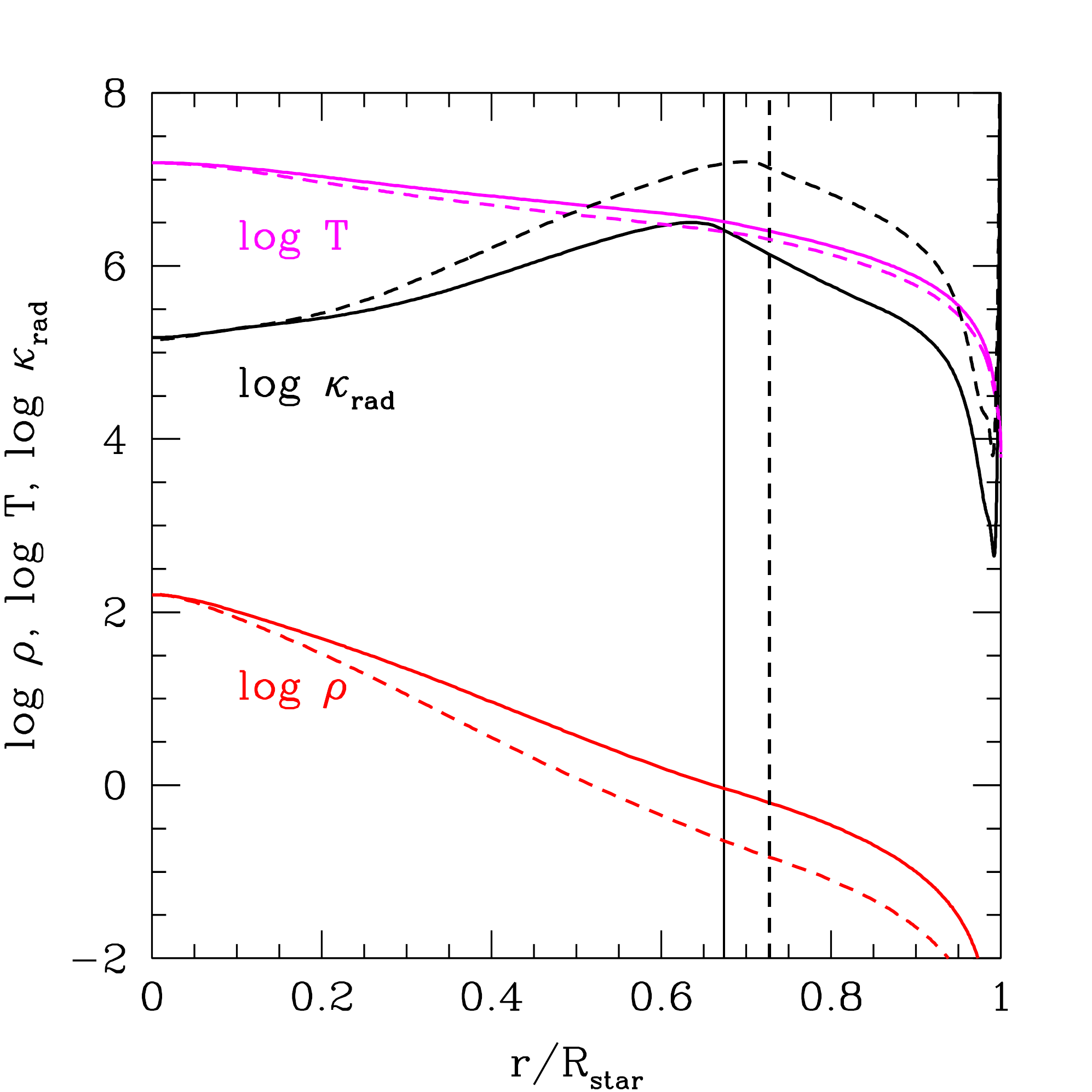}
\vspace{-0cm}
   \caption{Internal structure of the reference  model (solid lines) used as initial model for the 2D simulations compared to a  model for the current Sun (dashed lines). The figure shows the density $\rho$ (in g cm$^{-3}$), temperature T (in K) and radiative diffusivity $\kappa_{\rm rad} = \chi/ (\rho c_P)$ (in cm$^2$ s$^{-1}$), with $c_P$ the specific heat at constant pressure. The Schwarzschild boundary is indicated by the vertical solid line for the reference model and the vertical dashed line for the Sun-like model. }
 \label{struc_fig}
\end{figure}

The multi-D simulations require as initial input a radial profile of density and internal energy; for this work, those initial profiles are provided by the one-dimensional Lyon stellar evolution code \cp{baraffe91, baraffe98}, using the same opacities and equation of state as implemented in MUSIC. 
We have chosen stellar interior structures close to the Sun's structure, that is a  solar mass star on the Main Sequence with a convective envelope covering $\sim$ 30\% of the stellar radius. Our motivation is to use initial structures as close as possible to realistic stellar interior structures, as done in our previous studies \cp[see][]{pratt16, pratt17}. But some caution is needed to construct an initial model using a stellar evolution code if the goal is to test the impact of artificially enhanced luminosity and radiative diffusivity. In a typical solar model calculated with the Mixing Length Theory and a mixing length  $l_{\rm mix} = 1.9 H_P$, the superadiabaticity (see definition below)
in the bulk of the convective zone is very small, typically $<< 10^{-4}$, but in the outer convective zones the superadiabaticity is high, typically $> 10^{-2}$. 
We recall that the superadiabaticity is defined as ($\nabla - \nabla_{\rm ad}$) with $\nabla = {{\rm d} \log T \over {\rm d} \log P}$ the temperature gradient and $\nabla_{\rm ad} = {{\rm d} \log T \over {\rm d} \log P}|_S$ the adiabatic gradient. The Schwarzschild boundary is defined as the transition layer between convective instability ($\nabla > \nabla_{\rm ad}$) and stability ($\nabla < \nabla_{\rm ad}$).
The outer structure is thus very sensitive to any change of the opacity (and thus of the thermal diffusivity) and of the luminosity, as such changes will modify the superadiabaticity and thus the temperature stratification. Therefore, in order to avoid a readjustment of the model structure when starting a hydrodynamical simulation using MUSIC with enhanced luminosity and  thermal diffusivity, the profile of the stellar structure model must be adiabatic. We have thus constructed an artificial solar-like model with our stellar evolution code enforcing a very small superadiabaticity ($< 10^{-8}$) throughout the convective zone. In this case, an increase in the luminosity and of the radiative diffusivity (or a decrease of the opacity by the same factor) has no impact on the model structure (in terms of density and temperature radial profiles). This yields  a reference initial model slightly more compact and hotter than a model for the current Sun (that is a 1 $\msol$ model with the luminosity and radius of the current Sun) calculated with an initial helium abundance of 0.28, metallicity $Z=0.02$ and $l_{\rm mix} = 1.9 H_P$, as shown in Fig. \ref{struc_fig}. 

\begin{table}[t]
   \caption{Properties of the initial reference model: mass, luminosity, radius, depth of
   the convective envelope and pressure scale height at the Schwarzschild boundary.}
   \label{tab1}
   \centering
   \begin{tabular}{c c c c c } 
     \hline \hline
     $M/\msol$ &  $L_{\rm star}/L_\odot^{(a)}$ & $R_{\rm star}$ (cm) &  $r_{\rm conv}/R_{\rm star}$ & $H_{P,{\rm CB}}$ (cm) \\
      \hline
      1 &  1.07 & 5.64491 $\times$ 10$^{10}$ & 0.6734 & 5.086 $\times$ 10$^{9}$ \\
      \hline
   \end{tabular}
   \tablefoot{
      \tablefoottext{a}{We use $L_\odot = 3.839 \times 10^{33}$ erg/s.} 
   }
\end{table}

The properties of the reference model are displayed in Table \ref{tab1}.
We stress that starting numerical hydrodynamic simulations from realistic stellar structures with artificially enhanced luminosities and thermal diffusivities requires the superadiabaticity in the convective zone of the initial model to be extremely small. 
If not, hydrodynamical models will relax towards a structure that departs from the initial 1D structure,  and the larger the luminosity enhancement factor, the higher the departure because of the shorter thermal relaxation timescale.  Consequently, the conclusions one may derive by comparing simulations with different luminosity enhancement factors will be affected by the additional impact of having different structures. This would be equivalent to comparing different stellar models with different luminosities.

\subsection{Spherical-shell geometry and boundary conditions}
We perform two-dimensional simulations in a spherical shell using spherical coordinates, namely  $r$ the radius and $\theta$ the polar angle, and assuming azimuthal symmetry in the $\phi$-direction. The inner radius $r_{\rm in}$ is defined at 0.4 $R_{\rm star}$ and the outer radius at $r_{\rm out}$=0.9 $R_{\rm star}$. The angular extent ranges from $\theta = 0 ^\circ$ to $\theta = 180 ^\circ$, including the full hemisphere. We use a uniform grid resolution of $r \times \theta$ = 512 $\times$ 512 cells. This provides a good resolution of the pressure scale height at the Schwarzschild boundary $H_{P,{\rm CB}}/\Delta r \sim 92$, with $\Delta r$
=5.5 10$^7$ cm (550 km)  the radial grid spacing. In the $\theta$ direction, the typical size of a cell is 2300 km. The choice of the resolution in the $\theta$ direction is set by the requirement to preserve a good aspect ratio of the grid cells on the whole domain on a spherical grid.
We note that increasing the resolution by a factor two (1024 $\times$ 1024) does not change our results and conclusions (see discussion in \S \ref{conclusion}). 
In terms of boundary conditions,  we impose a constant radial derivative on the density on the inner and outer radial boundaries, as discussed in \ct{pratt16}.
For the reference model ref,  the energy flux at the inner and outer radial boundaries are set to the value of the energy flux at that radius in the one-dimensional stellar evolution model. For the artificially boosted simulations, the energy flux, and equivalently the luminosity,  at the boundaries is multiplied by an enhancement factor, and the Rosseland mean opacities $\kappa$ in MUSIC are decreased by the same factor.  In this work we analyse the impact of enhancing the luminosity and thermal diffusivity by factors 10, 10$^2$ and 10$^4$. 
 Larger enhancement factors (up to $10^6$) can be found in previous works \cp[e.g.][]{rogers06, hotta17}, but our selected range and values are appropriate for our general purpose, namely exploring the impact of artificially increasing the luminosity on the properties of overshooting and of internal waves.
In velocity, we impose on the radial boundaries non-penetrative condition for the radial velocity and stress-free boundary condition for the angular velocity. At the boundaries in $\theta$, because of the extension of the angular domain to the ``poles", we use reflective boundary conditions for the density and energy, meaning that they are mirrored at the boundary.  We adopt a stress free boundary condition for the radial velocity and a reflecting boundary condition for the angular velocity, to ensure it is equal to zero at the boundary.

\section{Results: Average dynamics and fluxes}

\subsection{General properties}
\label{dynamics}

\begin{table}[t]
   \caption{Summary of the 2D simulations.}
   \label{tab2}
   \centering
   \begin{tabular}{l c c c c c} 
     \hline \hline
     Model &  $L/L_{\rm star}$ & $\tau_{\rm conv}^{(a)}$ (s) &  $N_{\rm conv}^{(b)}$ &  $t_{\rm steady}^{(c)}$ (s) & $t_{\rm sim}^{(d)}$ (s)  \\
      \hline
      ref &  1 & 8 $\times 10^5$ & 565 & 1.3 $\times 10^8$ & 6.1 $\times 10^8$ \\
       boost1d1 &  10$^1$ & 3.6 $\times 10^5$ & 375 & 3 $\times 10^7$ & 1.72 $\times 10^8$ \\
        boost1d2 &  10$^2$ &   1.7 $\times 10^5$ & 450 & 2.44  $\times 10^7$ & 9.9  $\times 10^7$\\
        boost1d4 &  10$^4$ & 3.5 $\times 10^4$& 530 & $1.5 \times 10^6$ & $2 \times 10^7$\\
      \hline
   \end{tabular}
 \tablefoot{
\tablefoottext{a}{Convective turnover time. See \S \ref{dynamics} for its definition.} 
 \tablefoottext{b}{Number of convective turnover times covered by the simulation once  steady state convection is reached.} 
  \tablefoottext{c}{Physical time to reach a  steady state for convection.}
  \tablefoottext{d}{Total physical runtime of the simulation.} 
    }
\end{table}
\begin{figure}[h!]
\vspace{-1cm}
\includegraphics[height=10cm,width=8cm]{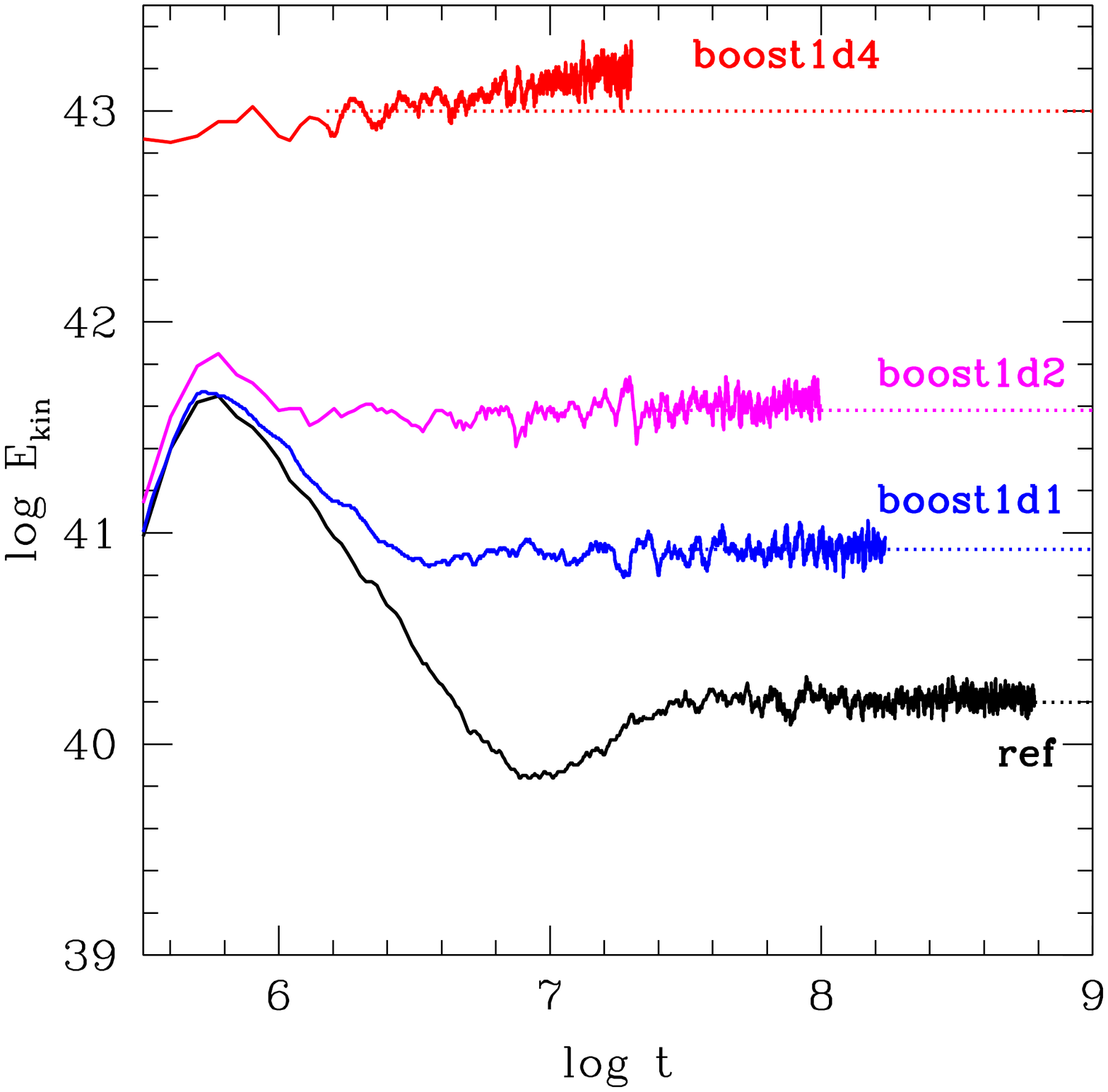}
\vspace{-0.5cm}
   \caption{Evolution of the total kinetic energy (in erg) as a function of time (in s) of the simulations described in Tab. \ref{tab2}. The dotted lines correspond to the value of the total kinetic energy at the beginning of the steady state for convection.}
 \label{ekin_fig}
\end{figure}

Our four simulations are summarized in Table \ref{tab2}. 
Figure \ref{ekin_fig} shows the time evolution of the total kinetic energy for the four models over the entire spherical volume
$E_{\rm kin} = \int_V {1 \over 2} \rho \velvec^2 {\rm d}V$  with $\rho$ the density, $\velvec$  the total velocity and ${\rm d}V$ the volume element. After a relaxation phase characterised by the propagation of strong acoustic waves and the onset of convection, $E_{\rm kin}$ reaches a plateau which characterises the  steady state for the convection. The time $t_{\rm steady}$ to reach this state is indicated in Table \ref{tab2} for each model. 
Our numerical simulations are not thermally relaxed. Achieving thermal relaxation is a well-known challenge for  global hydrodynamical simulations of convection based on realistic stellar interior structures \cp[e.g.][]{meakin07,horst20, higl21}. The global thermal relaxation or Kelvin-Helmholtz timescale of the initial stellar structure used for our simulations is given by $\tau_{\rm th} \sim G M^2/(RL) \sim 4 \times 10^6$ yr. Our total simulation times (see Table 2) remain much shorter than $\tau_{\rm th}$, even for the most boosted model analysed here and for which $\tau_{\rm th} \sim 4 \times 10^2$ yr. By specifically choosing as reference model a stellar model with no enhancement factor for the luminosity, this is unavoidable. As a consequence all these simulations are expected to maintain a secular drift. 
For the ref, boost1d1 and boost1d2 models, the drift is so slow that calculating statistical data and thermal properties during this very slowly changing transitional state is sensible.  However Fig. \ref{ekin_fig} shows that the most boosted model is so far out of balance that it is continuously evolving. As will be discussed in sections \ref{statistics}-\ref{fluctuations}, time averages for this model have limited meaning.

\begin{figure}[h!]
\vspace{-0.5cm}
\includegraphics[height=11cm,width=9cm]{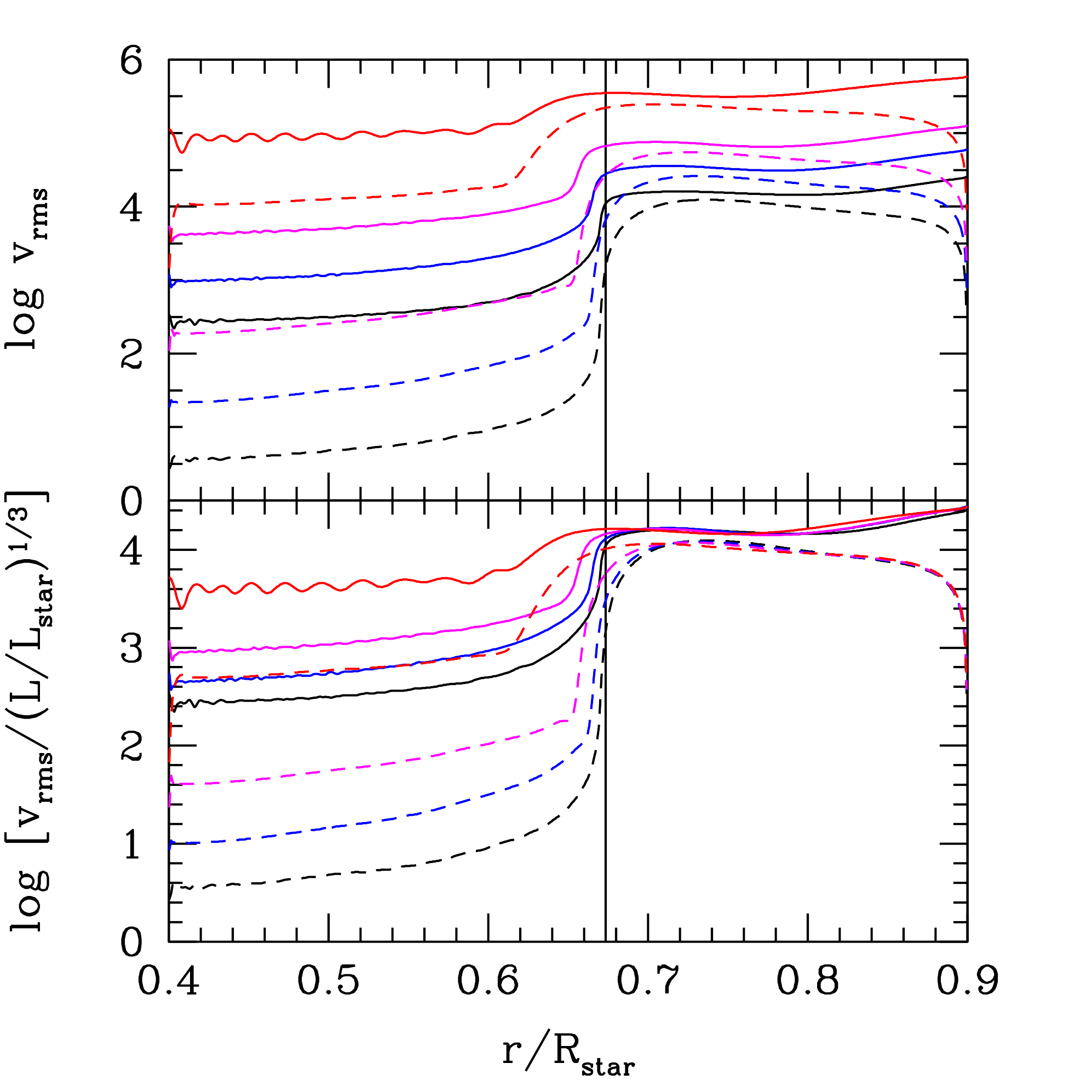}
\vspace{-0cm}
   \caption{Top panel: Radial profile of the time averaged rms velocity (solid lines) and rms radial velocity (dashed lines). Velocities are in cm/s. The curves from bottom to top correspond to the four models: ref (black), boost1d1 (blue), boost1d2 (magenta) and boost1d4 (red), respectively. Bottom panel: velocities scaled by the enhancement factor $(L/L_{\rm star})^{1/3}$. The convective boundary corresponding to the Schwarzschild boundary from the 1D initial model is indicated by the vertical solid line.}
 \label{vrms_fig}
\end{figure}

In the following, time averages are denoted by $\langle \rangle_t$ and calculated between $t_{\rm steady}$ and $t_{\rm sim}$ (see values in Table \ref{tab2}), i.e.  for any quantity $X$ we define:
\begin{equation}
\big \langle  X \big \rangle_t = {1 \over (t_{\rm sim} - t_{\rm steady})} \int_{t_{\rm steady}}^{t_{\rm sim}} X {\rm d}t.
\end{equation} 

We estimate a global convective turnover time $\tau_{\rm conv}$ based on the rms velocity $ \vel_\mathsf{rms}(r,t)$ at radius $r$ and time $t$, which characterises a bulk convective velocity:

\begin{equation}
\tau_{\rm conv} =  \Big \langle {{ \int_{r_{\rm conv}}^{r_{\rm out}} {{\rm d}r \over \vel_\mathsf{rms}(r,t)  }  }  }\Big \rangle_t,
\end{equation}
where the rms velocity is given by
\begin{equation}
\vel_\mathsf{rms}(r,t) = \sqrt {\langle \velvec^2(r,\theta,t) \rangle_\theta}, \label{eq_rms}
\end{equation}
with $\velvec^2 = \velvec_r^2 + \velvec_{\theta}^2$, $\velvec_r$ and $\velvec_{\theta}$ being the radial and angular velocities, respectively.  $\langle \rangle_\theta$ denotes a volume-weighted average in the angular ($\theta$) direction and is defined for any quantity X as
\begin{equation}
\big \langle X(r,\theta,t) \big \rangle_\theta = {\int_\theta  X(r,\theta,t) {\rm d}V(r,\theta) \over \int_\theta {\rm d}V(r,\theta)}.
\end{equation}
\noindent
The values of $\tau_{\rm conv}$ are provided in Table \ref{tab2} and the corresponding rms velocity profiles are shown in the top panel of Fig. \ref{vrms_fig}, for reference. The turnover time follows a scaling $\tau_{\rm conv} \propto L^{-1/3}$, as expected from the
scaling of $ \vel_\mathsf{rms}$ with luminosity $ \vel_\mathsf{rms} \propto L^{1/3}$ based on theoretical arguments from the mixing-length theory \cp{biermann32} and confirmed by many hydrodynamical simulations \cp[e.g.][]{porter00, viallet13, jones17, Edelmann19, andrassy20}. Our simulations  reproduce  this scaling as illustrated in the bottom panel of Fig. \ref{vrms_fig} which displays the rms  velocity and rms radial velocity for the four models. The rms velocities observed in the stably stratified region are due to the propagation of internal waves excited by the convective motions and penetrative plumes at the convective boundary.  Interestingly, the amplitude of these waves  also follow a scaling with the luminosity. We find that the rms velocities follow very closely a scaling  $ \vel_\mathsf{rms} \propto L^{0.61}$ for all models and for the radial velocity, the scaling is close to $ \vel_{r,\mathsf{rms}} \propto L^{0.86}$. Analysis of the internal waves is beyond the scope of this work and will be presented in a follow-up study. Our focus in the present study is on the process of overshooting. Figure \ref{vrms_fig} clearly shows deeper penetration of the convective motions beyond the convective boundary with increasing energy input. This is predicted by analytic models of overshooting \cp[e.g.][]{zahn91, rempel04} and found by numerical simulations in the literature  \cp[e.g.][]{hotta17, kapyla19}.  In the following section (see \S \ref{statistics}), we will quantitatively estimate  overshooting depths based on our approach developed in \ct{pratt17}. 
This approach relies on high order statistics of energy fluxes, such as the kinetic energy and heat fluxes, in contrast to the previous estimates which use only average quantities \cp[e.g.][]{hurlburt94, brummell02, rogers06, kapyla17}.

\begin{figure}[h!]
\vspace{-2cm}
\includegraphics[height=12cm,width=9cm]{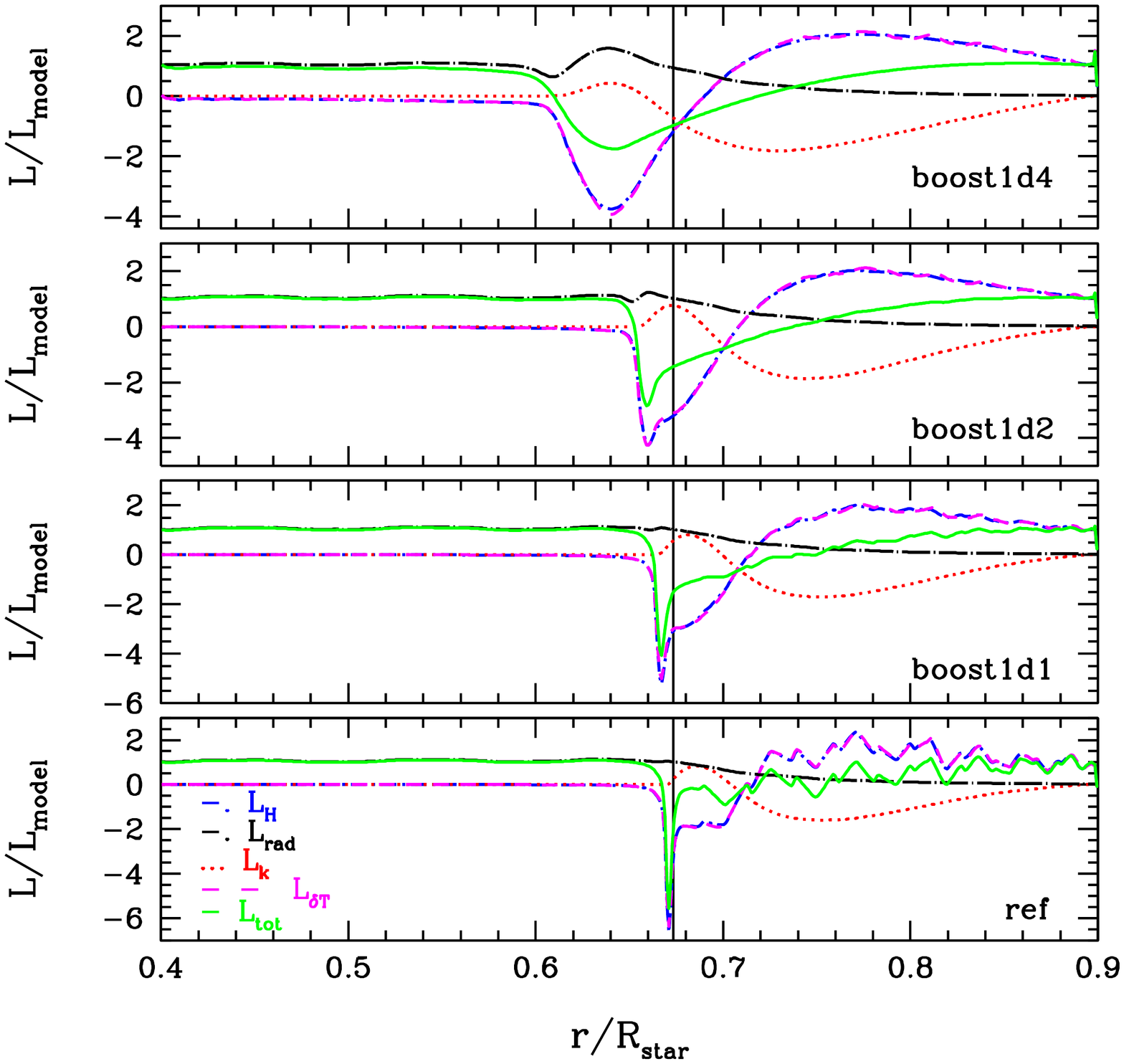}
\vspace{-2cm}
   \caption{Radial profile of luminosities for the four models as indicated in each panel, corresponding to various fluxes:  vertical enthalpy flux ($L_{\rm h}$, dash-dot, blue),  radiative flux ($L_{\rm rad}$, long dash-dot, black), vertical kinetic energy flux ($L_{\rm k}$, dot, red), convective heat flux ($L_{\rm \delta T}$, dash, magenta). The total luminosity $L_{\rm tot}=L_{\rm h} + L_{\rm rad} + L_{\rm k}$ (solid, green) is also displayed. Luminosities are divided by the luminosity of the corresponding model. The convective boundary corresponding to the Schwarzschild boundary from the 1D initial model is indicated by the vertical solid line.}
 \label{flux_fig}
\end{figure}
An inspection of averaged fluxes is  still interesting to further illustrate the dependence of penetration on luminosity.  At each radius $r$, we define a mean (time average and volume-weighted average in the $\theta$ direction) vertical kinetic energy flux $F_{\rm k}$,  enthalpy flux $F_{\rm h}$ or equivalently    convective heat flux $F_{\rm \delta T}$  by:
\begin{eqnarray}
\mathbf F_{\rm k} &=&\Big \langle { \big \langle {1 \over 2} \rho \velvec^2 \velvec_r \big \rangle_{\theta}} \Big \rangle_t, \label{eqk}\\
\mathbf F_{\rm h} &=& \Big \langle {\big \langle H \rho \velvec_r \big \rangle_\theta - \big \langle H \big \rangle_\theta  \big \langle \rho \velvec_r \big \rangle_\theta \Big \rangle_t, \label{eqh}}\\ 
\mathbf F_{\rm \delta T} &= & \Big \langle {\big \langle \rho c_P (\delta T) \velvec_r \big \rangle_\theta} \Big \rangle_t, \label{eqdt}
 \end{eqnarray}
where $H = e +P/\rho$ is the specific enthalpy and the second term in the r.h.s. of Eq. (\ref{eqh}) subtracts any non-zero mean vertical mass flux \cp{freytag96}. The temperature fluctuation $\delta T$ for each grid-cell and time $t$ is defined as
\begin{equation}
\delta T(r,\theta,t) = T(r,\theta,t) - \Big \langle {\big \langle T(r,\theta,t)\big \rangle_\theta } \Big \rangle_t.
\label{eqdelta}
\end{equation}

\begin{figure}[h!]
\vspace{0cm}
\includegraphics[height=9cm,width=9cm]{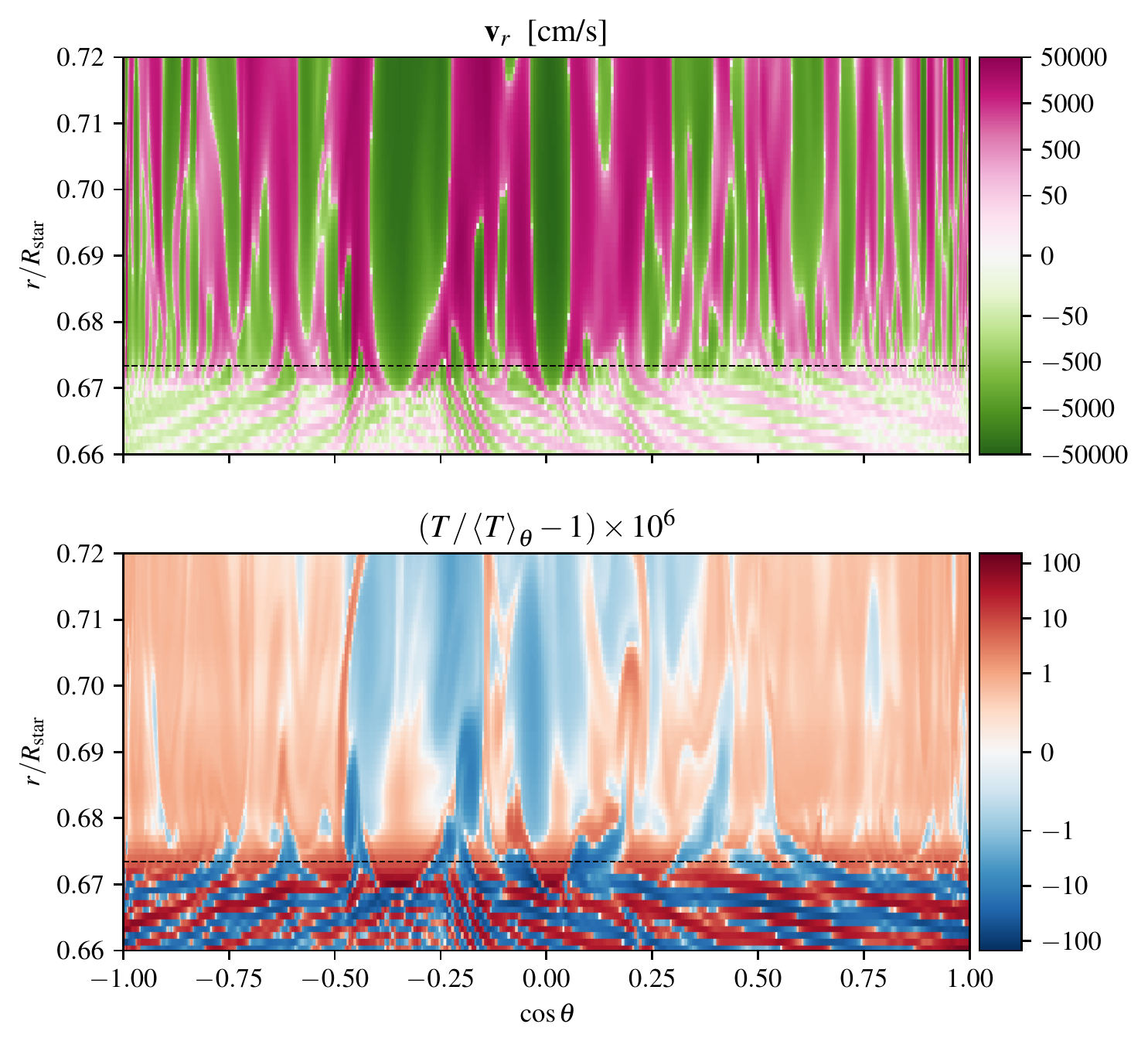}
\vspace{0cm}
   \caption{Visualisation of the radial velocity and the temperature fluctuations close to the convective boundary (horizontal black line) for the ref model at a given time. The x-axis represents the co-latitude (in terms of $\cos \theta$).}
 \label{snapshot_fig}
\end{figure}

The luminosities $L=4 \pi r^2 F$ corresponding to these fluxes are displayed in Fig. \ref{flux_fig} for the four models. Note that in the literature, the enthalpy flux is calculated using either  Eq. (\ref{eqh}) or Eq. (\ref{eqdt}). Both expressions give similar results as shown by the excellent agreement between these two quantities in Fig. \ref{flux_fig}.
We show in addition the mean radiative flux $F_{\rm rad}$ 
given by:
\begin{equation}
\mathbf F_{\rm rad} = \Big \langle {\big \langle - \chi \mathbf \nabla T\big \rangle_\theta} \Big \rangle_t,
\end{equation}
with $\nabla T = {\partial T \over \partial r}$.
The layer below the convective boundary where convective penetration proceeds can be characterised by the negative peak of the enthalpy flux \cp{hurlburt86, muthsam95, brun11, pratt17, korre19, kapyla19}. Convective downflows transport low entropy (cool material) from the stellar surface down to the bottom of the convective zone. Inspection of the temperature fluctuations indeed indicates that downflows are characterised by negative $\delta T$ and upflows by positive $\delta T$  in the convective zone (see Fig. \ref{snapshot_fig} and \S \ref{fluctuations}).  But when the downflows cross the convective boundary and penetrate the stably stratified medium, they are adiabatically compressed and therefore get hotter (positive $\delta T$) and less dense (negative density fluctuation $\delta \rho $) than the subadiabatically stratified  environment. Upward flows have the reverse pattern so both flow types contribute to a negative enthalpy flux in the layer where the bulk of the convective plumes penetrate.  

The properties of the flow and of the temperature fluctuations close to the convective boundary are shown in Fig. \ref{snapshot_fig} at a given time for the ref model, illustrating the patterns above described which are common to all models. The figure shows that the present simulations can capture the front of plumes penetrating across the convective boundary and the waves excited and propagating below it, even if our resolution does not allow to fully resolve their detailed structure.
The behaviour of the enthalpy flux in the penetration layer above described is a known feature and has already been discussed in detail in the literature \cp[e.g.][]{hurlburt86, muthsam95}. However, less attention has been given to the impact on the thermal profile and more specifically to the feedback between the radiative flux and the temperature profile in this layer. This is the focus of the next section.

\subsection{Modification of the thermal background}
\label{heating}

 The position of the negative peak of the enthalpy flux  moves deeper inward as the luminosity of the model increases, confirming the larger extension of the overshooting depth with input flux already illustrated with the analysis of the rms velocities (see Fig. \ref{vrms_fig}). We can clearly see for the boosted models that the position of the negative peak of the enthalpy flux corresponds  to the position of a peak of the radiative flux. A peak of the radiative flux also exists for the ref model, although much less pronounced. 
This feature has also been observed in previous numerical simulations \cp{rogers06, brun11, brun17, kapyla19, cai20}. 
In \ct{brun11, brun17}, the authors explain that ideally the system would compensate for the negative heat flux, considered as a flux deficit, and adjust to a new equilibrium by modifying the background thermal stratification in a thermal timescale. In order to achieve satisfactory flux balance, \ct{brun11, brun17} arbitrarily increase the radiative diffusivity near the base of the convective zone, in order to  accelerate the thermal relaxation process. This thermal adjustment yields a peak of the radiative flux in order to compensate  for the flux deficit near the base of the convective zone. 

\begin{figure}[h!]
\vspace{-0.5cm}
\includegraphics[height=11cm,width=8cm]{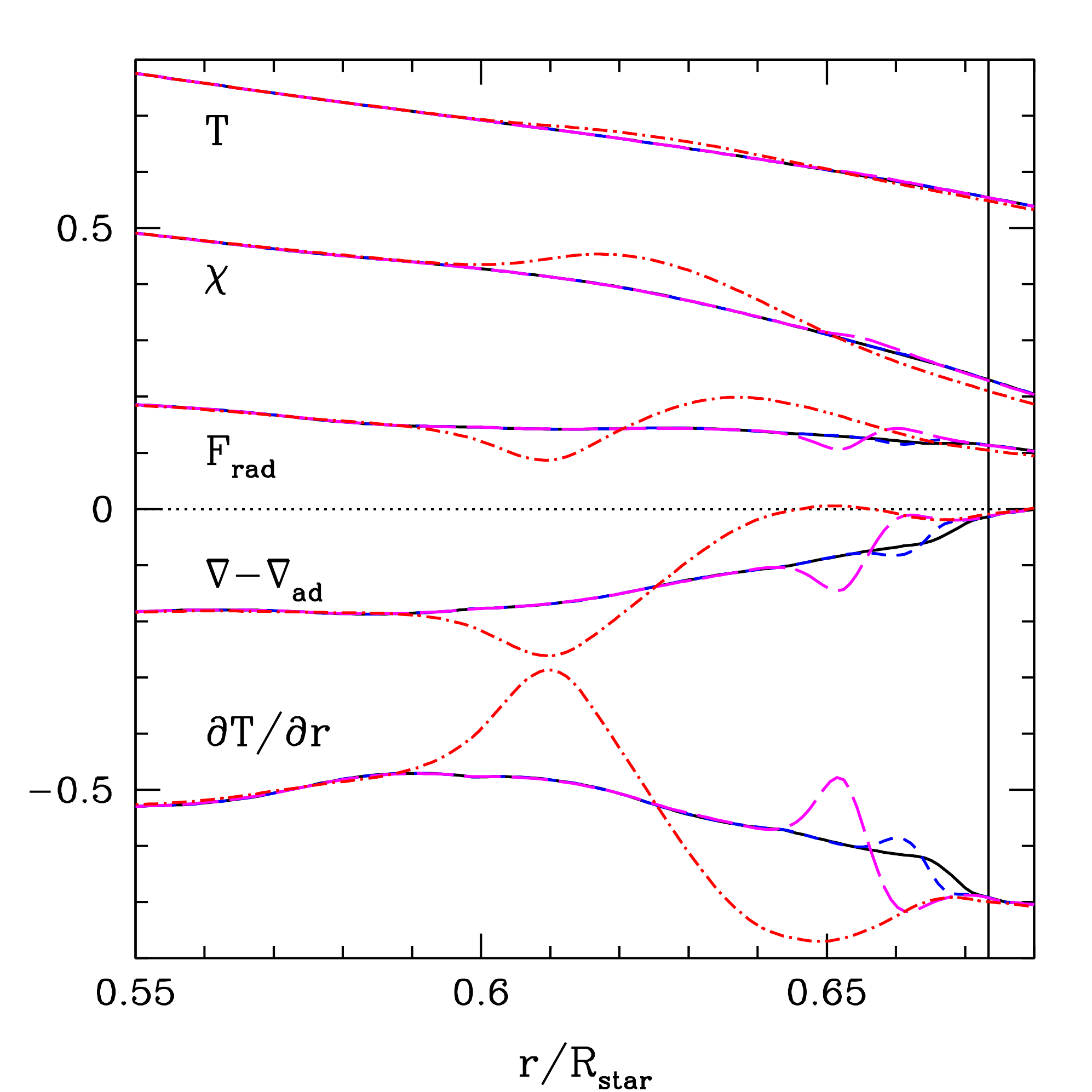}
\vspace{-0cm}
   \caption{Radial profile of time and space averages of various quantities close to the convective boundary. From top to bottom:  the temperature $T$ (in units of 6 $\times 10^6$ K); the radiative conductivity $\chi$ divided by the luminosity enhancement factor of the model and by the constant 4 $\times 10^{15}$; the radiative flux divided by the luminosity enhancement factor of the model and by the constant 2 $\times 10^{12}$; the subadiabaticity $\nabla - \nabla_{\rm ad}$; the radial temperature gradient $\partial T / \partial r$  multiplied by the constant 2.8 $\times 10^{3}$. The coloured curves correspond to the four models: ref (black), boost1d1 (blue), boost1d2 (magenta) and boost1d4 (red). The convective boundary corresponding to the Schwarzschild boundary from the 1D initial model is indicated by the vertical solid line.}
 \label{temp_fig}
\end{figure}

In the  simulations of \ct{rogers06} devoted to a solar model, similar results are found with the temperature gradient in the penetration region becoming slightly steeper, a feature that is interpreted in the same way as  \ct{brun11, brun17}, namely that the upward diffusive heat flux  increases to compensate for the negative convective heat flux in that region. 
\ct{rogers06} also find that the region heats up because convective motions are continually pumping heat into this region and the timescale for this heat transfer is much shorter than the diffusive timescale. This local heating causes a steeper temperature profile and a less subadiabatic temperature gradient, i.e. $|\nabla - \nabla_{\rm ad}|$ decreases, immediately below the convective boundary. Immediately below this region, 
 the temperature profile is slightly flatter (hotter) and the temperature gradient becomes more subadiabatic, i.e. $|\nabla - \nabla_{\rm ad}|$ increases \cp[see Fig. 3 in][]{rogers06}. Note that \ct{korre19} also find that the fluid motions affect the thermal stratification at the base of the convective zone in the penetration region. But since  \ct{korre19} use the Boussinesq approximation, it is not clear how the modification of the thermal background is accounted for in such framework. Finally, in 2D simulations for convective core overshooting, \ct{higl21} note also a modification of the temperature stratification in the penetration region compared to the initial 1D model. Despite the fact that their simulations are not thermally relaxed, they suggest that  penetration will very likely impact the final thermal background in the overshooting region.

We find similar features (see Fig.  \ref{temp_fig}):  a slight heating of the layers below the position of the negative peak of the heat (enthalpy) flux. The peak and the non-monotonic profile of the radiative flux is a consequence of the slight change of the averaged thermal profile, which impacts the radiative conductivity $\chi$ and the radial temperature gradient ${\partial T \over \partial r}$. 

In order to understand the origin of the local heating, we examine the rate-of-strain tensor $\mathbf s$ which has the following components in 2D spherical coordinates:
\begin{equation}
s_{rr}    =  {\partial \vel_r \over \partial r}, \\
\end{equation}
\begin{equation}
s_{\theta \theta}   =   {1 \over r}{\partial \vel_\theta \over \partial \theta} + {\vel_r \over r}, \\
\end{equation}
\begin{equation}
s_{r \theta}  = s_{\theta r}   =   {1 \over 2} \left[ r {\partial \over \partial r} \left({\vel_\theta \over r}\right) + {1 \over r} {\partial \vel_r \over \partial \theta} \right].
\end{equation}

\begin{figure}[h!]
\vspace{0cm}
\includegraphics[height=11cm,width=8cm]{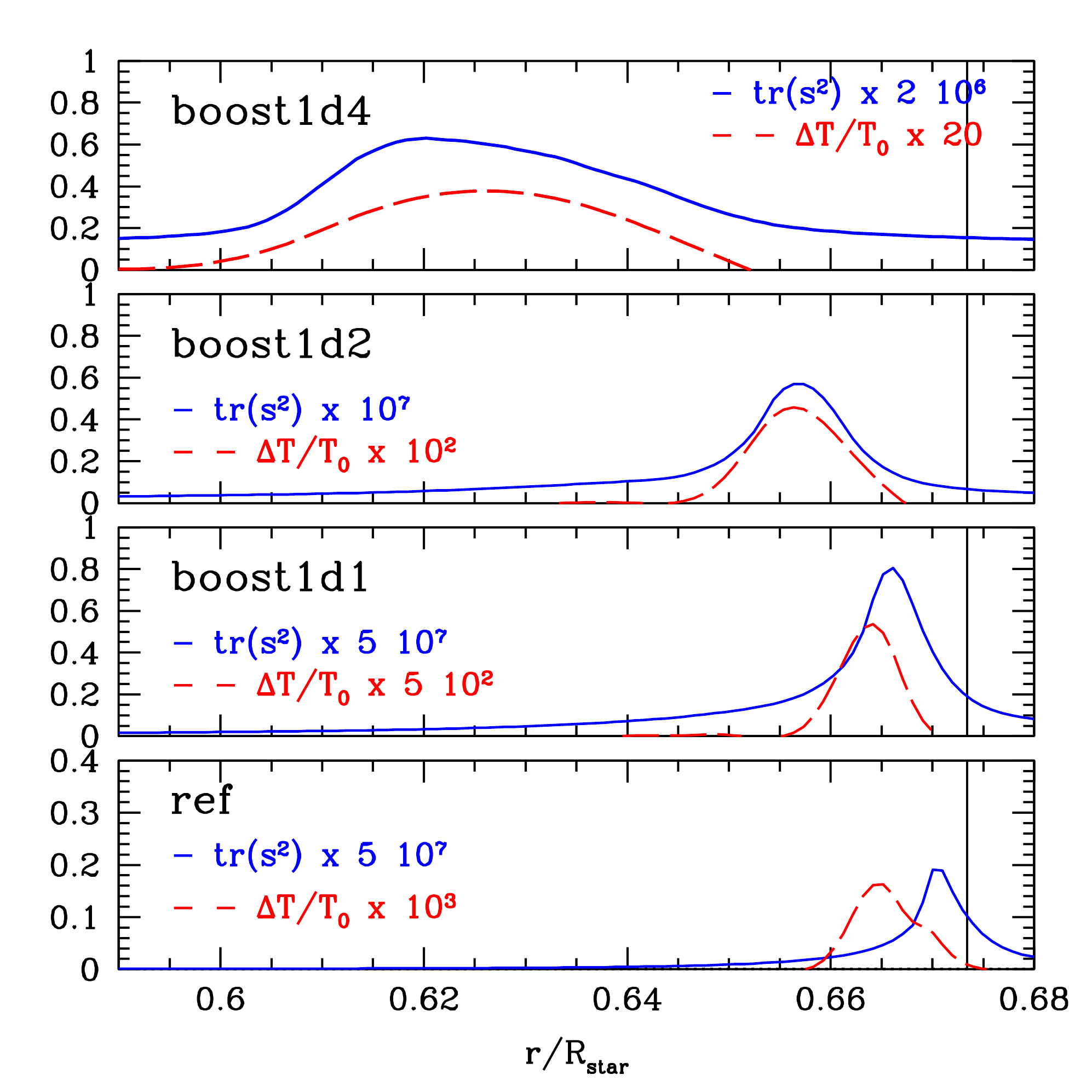}
\vspace{0cm}
   \caption{Radial profile of the time and space averages of the trace of  $\mathbf s^2$, $ {\rm tr} (\mathbf s^2) = s_{rr}^2 + s_{\theta \theta}^2 + 2 s_{r \theta}^2$ (solid blue lines) in the overshooting layer. The red dashed lines indicate the profile of the relative difference between the time/space average of the temperature $\big \langle  \langle T \rangle_\theta \big \rangle_t$ and the initial profile $T_0$.
 The convective boundary corresponding to the Schwarzschild boundary from the 1D initial model is indicated by the vertical solid line.}
 \label{trs2_fig}
\end{figure}

Figure \ref{trs2_fig} shows the trace of $\mathbf s^2$, which combines the contributions from compression and from shear, in the region below the convective boundary. Both contributions 
are due to the penetrative downflows that are compressed and braked by the stratification, inducing vertical and horizontal shears in this region. 
The region with the largest departure of the mean temperature profile compared to the initial profile coincides with the region with the largest trace of $\mathbf s^2$, suggesting that  compression and shear induce  local heating and thermal mixing (through mixing of hot material). The coincidence is not particularly obvious for the ref model, but there is an overlap between the two regions. We note that for this model the variation of temperature is very small, of the order of 0.01\%. 
The clear presence of these features which are intensified in the three  boosted models helps identifying them in the ref model. 

The additional heat injected by the downflows cannot be efficiently evacuated by radiative transport. 
Even in the  model boost1d4, despite the enhancement of the radiative diffusivity, radiative transport remains insufficient to transport the flux excess on the timescales considered in the present simulation. At the base of the convective boundary, the radiative diffusivity $\kappa_{\rm rad}$ of the reference model is $\sim 3 \times10^6$ cm$^2$ s$^{-1}$ (see Fig. \ref{struc_fig}). Over a characteristic length scale  $H_P$, the radiative timescale for the model boost1d4 $H_P^2/(\kappa_{\rm rad} \times 10^4)$  is $\sim 8 \times 10^8$s at the convective boundary, which is $\sim$ 40 times greater than the simulation time (see Tab. \ref{tab2}).
Heat thus accumulates and the structure is adapting in order to evacuate this heat excess.
The local modification of the thermal profile for the models boost1d2 and boost1d4, and the subsequent impact on the temperature gradient and the radiative flux, are clearly seen in Fig. \ref{temp_fig}. For the models ref and boost1d1 the features are much less pronounced, but are present. 


The local heating and the modification of the thermal background should proceed until the increase in radiative flux counterbalances the negative enthalpy flux and an equilibrium can be reached. The modification of the temperature gradient impacts the dynamics of the plumes. Larger subadiabaticity $|\nabla - \nabla_{\rm ad}|$  will provide a stronger resistance to penetrating motions, since the adiabatically compressed penetrating plume will have a larger temperature excess w.r.t. its surrounding
and will thus be more efficiently braked by buoyancy. This behaviour has been illustrated by studies analysing the sensitivity of the penetration process as a function of the so-called stability (or stiffness) parameter $S$,  which measures the ratio of subadiabaticity to superadiabaticity \cp{hurlburt94, brummell02}.  Conversely, smaller subadiabaticity reduces the braking effect and allows plumes to go deeper, broadening the overshooting layer and propagating the heating in deeper layers. The layers where $|\nabla - \nabla_{\rm ad}|$ increase w.r.t. to the initial profile (larger subadiabaticity) is a region which will be able to efficiently brake the most vigorous plumes. An equilibrium is reached when the modification of the thermal profile is sufficient to increase the radiative flux and to stop the inward progression of the overshooting layer and of the local heating. There is thus a complex connection between the strength of the plumes, their thermal properties  and the thermal background.  
This was noted by \ct{viallet13} when comparing the properties of convective penetration  in the simulations of a red giant  convective envelope and of an oxygen-burning shell.  They found a quasi-steady state in the red giant model in which non-adiabatic processes due to radiative transport can counter-balance the effects of turbulent entrainment and stop the growth in size of the convective region.

In contrast, our results show that for the most boosted model, the local heating of the thermal profile is continuously increasing with time, with the temperature gradient getting steeper and closer to the adiabatic gradient. The thermal profile continues heating without reaching an equilibrium state within the simulated time scales, and the temperature gradient may at some point become locally unstable, as can been seen in Fig. \ref{temp_fig} for the model boost1d4. We consider this as an extreme and artificial situation, due to combined effects, and interpret this behaviour as follows. The enhanced luminosity yields stronger penetrative plumes, which enhance the heating by compression and shears.  The modification of the thermal background is more pronounced and proceeds faster due to the enhanced thermal diffusivity (and thus smaller thermal relaxation time). It increases locally the radiative flux and the heating could stop if the radiative flux could evacuate the heat excess. But it also flattens the subadiabaticity, reducing the braking of penetrative plumes and the strongest plumes progress deeper, broadening the penetration region. This is a kind of runaway situation that may stop with a significant extension of the convectively unstable region to deeper levels, which seems unrealistic for a ``real'' stellar model. Note that even larger enhancement factors for the luminosity, with factors $> 10^6$, will produce even more unrealistic results with convective velocities in the outer part of the domain that would become close to the speed of sound.

An artificial modification of the heat conductivity in the penetration region, as done in \ct{brun11, brun17}, will obviously have a significant impact on the thermal profile, the temperature gradient and thus the dynamics of the penetrative plumes. Increasing the radiative transport efficiency in the overshooting region will not only accelerate the thermal relaxation process, but may also change the subtle balance  mentioned above and consequently the characteristic penetration depths. 
There is thus a subtle equilibrium between thermal background modification and plume dynamics, which will eventually determine the maximal depths of penetrative plumes (see \S \ref{statistics}).

Since our simulations are not thermally relaxed, it is an open question whether the effects that we describe above and  the modification of the thermal background that is linked to the bump of the radiative flux are transient and vanish once thermal relaxation is reached. 
But the bump of the radiative flux is also observed in thermally relaxed simulations \cp[e.g.][]{rogers06,kupka18,kapyla19}. Our experiments thus link the radiative flux bump to the modification of the local thermal background due to the penetrating downflows. They point the way towards a deeper understanding of the underlying physical processes responsible for this bump also observed in thermally relaxed state.
One has, however, to take with caution the direct application of the overshooting depths predicted by these simulations to ``real'' stars  since the final relaxed state for these simulations may have different properties from present non thermally relaxed states. This does not however preclude analysing the effect of luminosity and radiative diffusivity enhancement during the slowly evolving transient phase during which convection is considered to be in steady state, at least for the ref, boost1d1 and boost1d2 models. 

\section{Results: Statistics of convective penetration depths}
\label{statistics}

In order to determine the extent of convective penetration, we adopt the approach developed in \ct{pratt17} based on a statistical analysis of the depth of all convective plumes that penetrate below the convective boundary. We use two criteria to determine the depth of a penetrative plume at a given angle $\theta$  and time $t$. The first one is given by the first zero below the convective boundary of the vertical kinetic energy flux 
$\mathbf f_{\rm k}(r,\theta,t) =  {1 \over 2} \rho \velvec^2  \velvec_r  $ and the second one considers instead the vertical heat flux $\mathbf f_{\rm \delta T}(r,\theta,t) = \rho c_P (\delta T) \velvec_r$.

\ct{pratt17}  show that the statistical analysis of penetration depths calculated for each angular grid cell at each time step reveals a non-Gaussian probability distribution and the tail of the distribution, due to extreme events of plume penetration, plays a key role in defining the extent of the penetration layer. The width of the penetration layer, where mixing proceeds, should thus be better characterised by the maximal depths of plume penetration than by the average depth.  Additionally, while there is significant discrepancy between the penetration depth defined from the {\it averaged} vertical kinetic energy flux (e.g. based on Eq. (\ref {eqk})) and the {\it averaged} vertical heat flux (e.g. based on Eq.(\ref{eqdt})), \ct{pratt17} finds that the PDFs of convective penetration depths calculated from the local quantities $\mathbf f_{\rm k}$ and $\mathbf f_{\rm \delta T}$ are characteristically similar. 

\ct{pratt17} use extreme value theory  to derive the cumulative distribution function of the maximal penetration depth obtained from numerical simulations. This method allows to infer a diffusion coefficient describing the mixing driven by the convective plumes in the penetration layer. This diffusion coefficient can be used in stellar evolution codes \cp[see e.g.][]{baraffe17}. 
 The fit to the generalised extreme value distribution, which is used  to model the probability of maximal events, depends in particular on a location parameter $\mu$ which is approximately equivalent to a maximal penetration depth and provides  a simple overshooting length \cp{pratt17,pratt20}. 
Our goal presently is not to determine diffusion coefficients but to analyse the impact of enhanced luminosity and radiative diffusivity on penetration depths. 
We thus focus on the comparison and characterisation of the distribution of all penetration depths and of maximal depths.
We use the same criteria as \ct{pratt17}  based on the two fluxes $\mathbf f_{\rm k}$ and $\mathbf f_{\rm \delta T}$, respectively, to define the penetration depth of a plume.   At each time $t$, we calculate at each angle $\theta$ the penetration depth $r_0(\theta,t)$ of a plume (i.e. the radial position of the first zero of $\mathbf f_{\rm k}$ and $\mathbf f_{\rm \delta T}$ below $r_{\rm conv}$, the position of the convective boundary given by the position of the Schwarzschild boundary from the 1D initial model) and define the corresponding penetration length $l_0$
with respect to  $r_{\rm conv}$ by
\begin{equation}
l_0 (\theta,t) = r_{\rm conv} - r_0(\theta,t). 
\end{equation}
Note that with this definition the length $l_0$ is always positive.  The maximal penetration length $l_0^{\rm max}$  is defined by the maximum over all angles $\theta$:
\begin{equation}
 l_0^{\rm max}(t) = \mathsf{max}(l_0(\theta,t)).
 \end{equation}
 The simulation data are sampled at a fixed time interval, typically at a rate of $\sim \tau_{\rm conv}/10^2$. Larger time intervals would not allow to capture the fastest moving plumes entering the penetration layer. 

Following \ct{pratt17}, we consider two different layers. We define a first layer of characteristic length $l_{\rm bulk}$ where convective plumes frequently penetrate and with $l_{\rm bulk}$  given by the average of all lengths $l_0$. The second, deeper, layer is characterised by convective plumes that penetrate intermittently and its length $l_{\rm max}$ is given by the average of all $l_0^{\rm max}$, providing an effective width of the overshooting layer. In this approach, the overshooting length $l_{\rm max}$, thus defined, is expected to be close to the location parameter $\mu$ inferred from a more complete extreme value statistical analysis.
Table \ref{tabdepth} displays $l_{\rm bulk}$  and $l_{\rm max}$  in units of the total stellar radius and of the pressure scale height at the convective boundary for the four models.

\begin{figure}[h!]
\vspace{0cm}
\includegraphics[height=4.4cm,width=4.4cm]{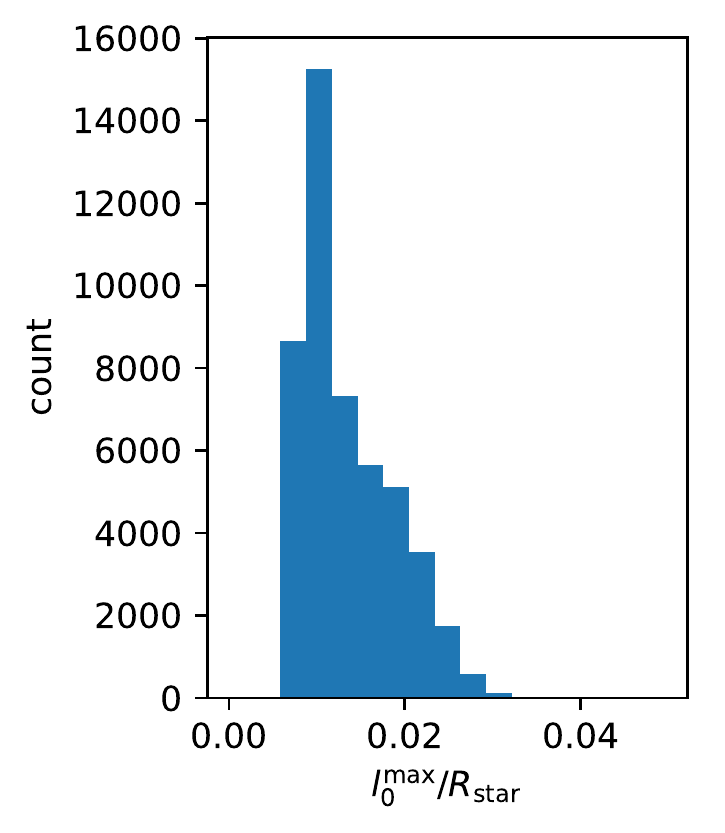}
\includegraphics[height=4.4cm,width=4.4cm]{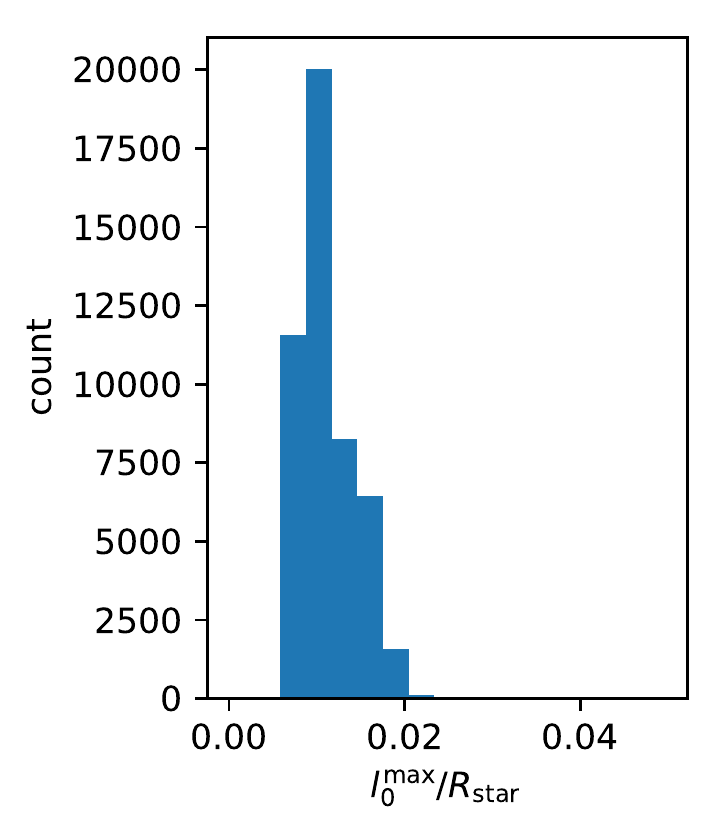}
\vspace{0cm}
   \caption{Histograms of maximal penetration lengths  $l_0^{\rm max}$, in units of the stellar radius $R_{\rm star}$, for the reference model ref, based on the vertical kinetic energy flux $\mathbf f_{\rm k}$ (left panel) and the vertical heat flux $\mathbf f_{\rm \delta T}$ (right panel).}
 \label{histkrad_fig}
\end{figure}

\begin{figure}[h!]
\vspace{0cm}
\includegraphics[height=4.4cm,width=4.4cm]{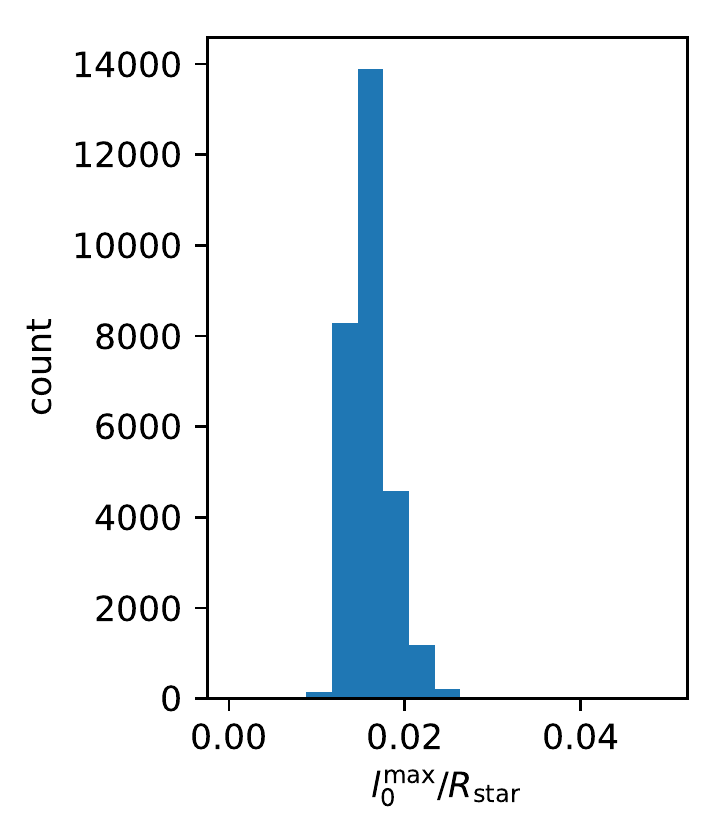}
\includegraphics[height=4.4cm,width=4.4cm]{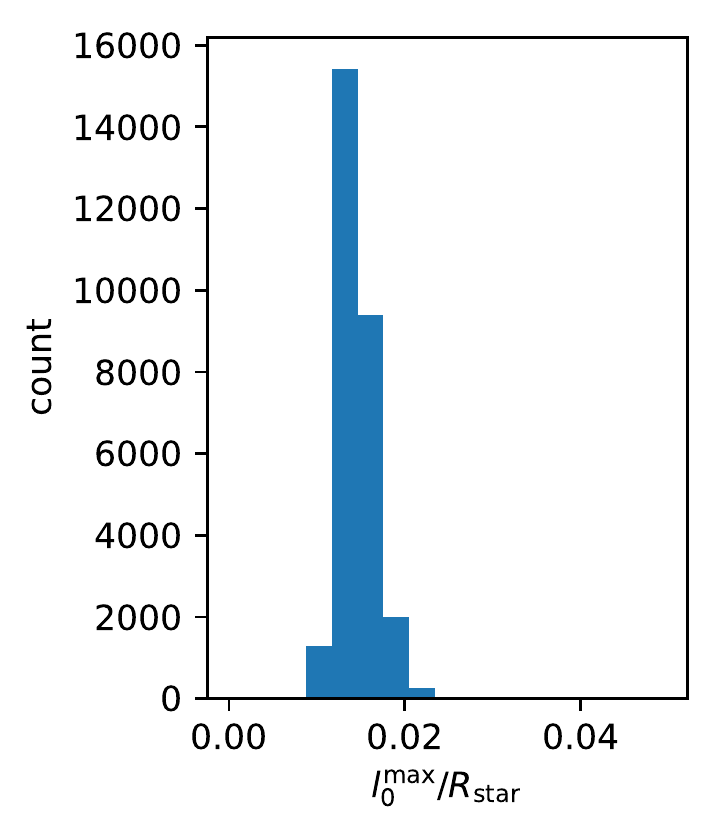}
\vspace{0cm}
   \caption{Same as Fig. \ref{histkrad_fig}, but for the model boost1d1.}
 \label{histkrad1d1_fig}
\end{figure}

\begin{figure}[h!]
\vspace{0cm}
\includegraphics[height=4.4cm,width=4.4cm]{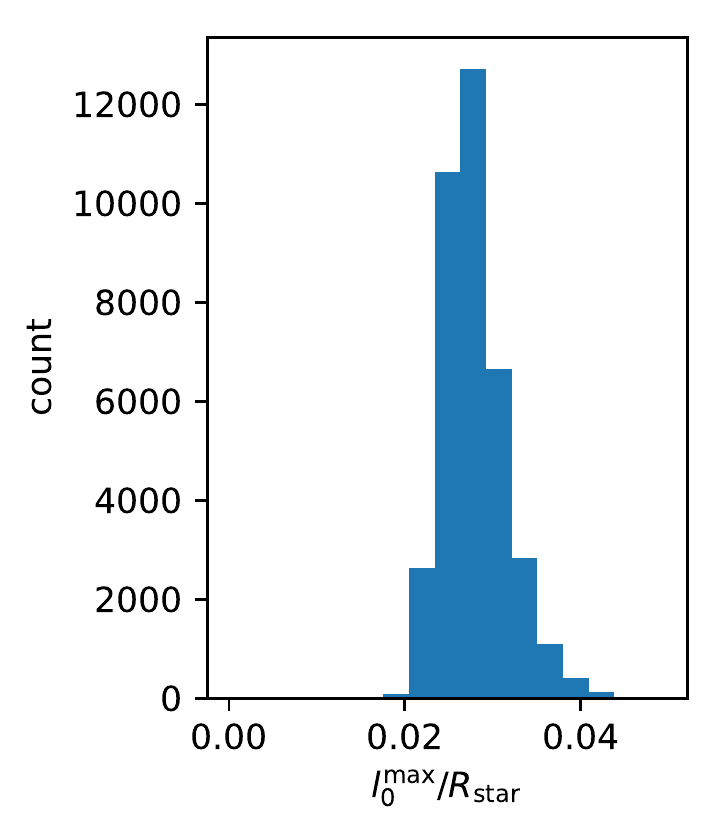}
\includegraphics[height=4.4cm,width=4.4cm]{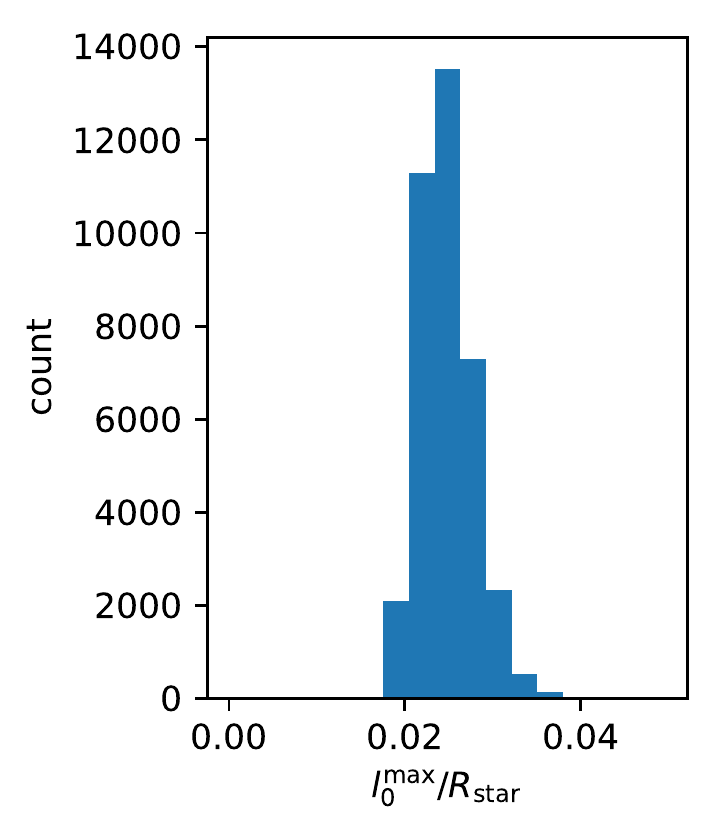}
\vspace{0cm}
   \caption{Same as Fig. \ref{histkrad_fig}, but for the model boost1d2.}
 \label{histkrad1d2_fig}
\end{figure}

\begin{figure}[h!]
\vspace{0cm}
\includegraphics[height=4.4cm,width=4.4cm]{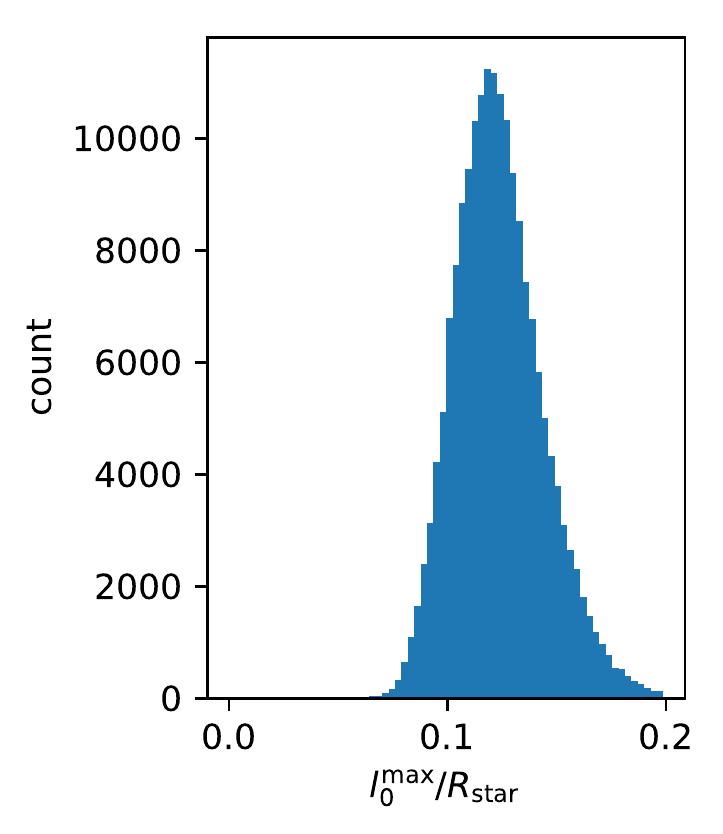}
\includegraphics[height=4.4cm,width=4.4cm]{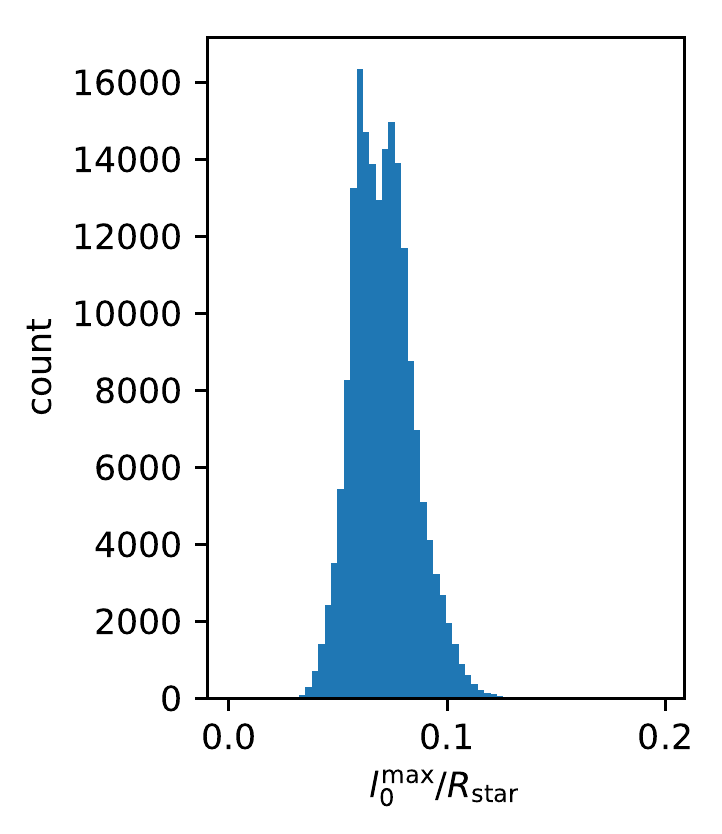}
\vspace{0cm}
   \caption{Same as Fig. \ref{histkrad_fig}, but for the model boost1d4. Note that the bins have the same width as in Figs. \ref{histkrad_fig}-\ref{histkrad1d2_fig}.}
 \label{histkrad1d4_fig}
\end{figure}

\begin{table*}[t]
   \caption{Characteristic lengths $l_{\rm bulk}$ and  $l_{\rm max}$, in units of the total stellar radius and of the pressure scale height at the convective boundary, for the four models considered in this study.}
   \label{tabdepth}
   \centering
   \begin{tabular}{l c c c c c c c c } 
     \hline \hline
     Model &  $l_{\rm bulk}(\mathbf f_{\rm k})/R_{\rm star}$ & $l_{\rm bulk}(\mathbf f_{\rm \delta T})/R_{\rm star}$  &  $l_{\rm max}(\mathbf f_{\rm k})/R_{\rm star}$  &  $l_{\rm max}(\mathbf f_{\rm \delta T})/R_{\rm star}$  & $l_{\rm bulk}(\mathbf f_{\rm k})/H_{P{\rm,CB}}$  &$l_{\rm bulk}(\mathbf f_{\rm \delta T})/H_{P{\rm,CB}}$   &  $l_{\rm max}(\mathbf f_{\rm k})/H_{P{\rm,CB}}$  &  $l_{\rm max}(\mathbf f_{\rm \delta T})/H_{P{\rm,CB}}$\\
      \hline 
      ref & 2.90 $\times 10 ^{-3}$ & 2.35 $\times 10 ^{-3}$ & 1.34 $\times 10 ^{-2}$ &1.12 $\times 10 ^{-2}$ & 3.22 $\times 10 ^{-2}$  &  2.62 $\times 10 ^{-2}$  & 0.149  & 0.124 \\
       boost1d1 & 5.80 $\times 10 ^{-3}$ &4.72 $\times 10 ^{-3}$ & 1.63 $\times 10 ^{-2}$ & 1.45 $\times 10 ^{-2}$ &6.44  $\times 10^{-2}$   &  5.24  $\times 10^{-2}$  & 0.181& 0.161 \\
        boost1d2 & 1.18  $\times 10 ^{-2}$ & 1.04 $\times 10 ^{-2}$  &2.80 $\times 10 ^{-2}$  & 2.47 $\times 10 ^{-2}$  &  0.131  & 0.115    &  0.311  & 0.275\\
        boost1d4 & 3.54 $\times 10 ^{-2}$ & 2.82 $\times 10 ^{-2}$ & 0.124 &   0.071 &  0.393  & 0.313 &  1.377 & 0.787 \\
      \hline
   \end{tabular}
\end{table*}

Figures \ref{histkrad_fig} - \ref{histkrad1d4_fig} show the distributions of maximal lengths based on the criterion for $\mathbf f_{\rm k}$ and $\mathbf f_{\rm \delta T}$, respectively, for our four models. For the models boost1d1 and boost1d2, the histograms of $l_0^{\rm max}$  derived from the vertical kinetic energy flux and  the vertical heat flux, respectively, are similar, confirming the findings of \ct{pratt17}. For these two models, we obtain $\sim$ 12 \% difference between the  mean value for $l_0^{\rm max}$  based on $\mathbf f_{\rm k}$, denoted $l_{\rm max}(\mathbf f_{\rm k})$, and the one based on $\mathbf f_{\rm \delta T}$, denoted $l_{\rm max}(\mathbf f_{\rm \delta T}$) (see Tab. \ref{tabdepth}). 
Note that when the same calculation is performed using only the first
half of both simulations, i.e. $N_{\rm conv} \sim 200$ for boost1d1 and  $N_{\rm conv} \sim 220$ for boost1d2, the difference in means between the two methods is $\sim$ 18\% in both cases.
Simulations over several hundred convective turnover times ($\simgt$ 400 $\times$ $\tau_{\rm conv}$) are thus necessary for the statistics to converge.
We also calculate the location parameter $\mu$  of the generalised extreme value distribution for penetration depths based on 
$\mathbf f_{\rm k}$. We find $\sim$ 1\% difference between $l_{\rm max}$ and $\mu$ for the model boost1d1 and  $\sim$ 5\% for the model boost1d2. This  confirms that our approach to estimate an overshooting length from the mean of all maximal lengths $l_0^{\rm max}$  is a good first approximation to the location parameter derived from the more complete statistical analysis based on extreme value theory. Note that this is valid if the goal is to perform a simple comparison between different models,  but the mean of all $l_0^{\rm max}$ will not reveal the  wealth of information on extreme plume events contained in the generalised extreme value distribution.

For the reference model, the shape of the histograms differ (see left and right panels of Fig. \ref{histkrad_fig}) and the difference between $l_{\rm max}(\mathbf f_{\rm k})$ and $l_{\rm max}(\mathbf f_{\rm \delta T})$
is $\sim$ 20 \%. But the two distributions become more similar as the simulation time
increases.
In particular the distribution based on $\mathbf f_{\rm k}$ is getting narrower with time and closer to the distribution based on $\mathbf f_{\rm \delta T}$. We stopped this simulation after $\sim$ 565 convective turnover times, considering that a difference of $\le$ 20\% between $l_{\rm max}(\mathbf f_{\rm k})$ and $l_{\rm max}(\mathbf f_{\rm \delta T})$ is satisfactory for our present purpose.

As expected from the results in \S \ref{heating}, the model boost1d4 shows a different behaviour compared to the less boosted models. 
This model does not reach a steady state since the thermal background in the overshooting region continues to evolve and the two penetration lengths $l_{\rm bulk}$ and $l_{\rm max}$ continue increasing, indicating that the penetration layer moves deeper inwards with time.
The plume depth analysis as performed here has to be taken with caution and is essentially illustrative.
The histograms and overshooting lengths based on $\mathbf f_{\rm k}$ and $\mathbf f_{\rm \delta T}$, respectively,  significantly differ, and the discrepancy does not decrease with time. 
The overshooting length predicted by the vertical kinetic energy flux is significantly larger than the one predicted by the vertical heat flux. 
This suggests that the velocity magnitude $\velvec^2$ and the temperature fluctuation ${\rm \delta T}$ react differently in this highly boosted model compared to the less boosted ones. 
Strong internal waves are generated at the convective boundary by the convective motions and by the penetrating plumes, as illustrated by the rms velocity of the boost1d4 model in the radiative zone (see  Fig. \ref{vrms_fig}). The large discrepancy between $l_{\rm max}(\mathbf f_{\rm k})$ and $l_{\rm max}(\mathbf f_{\rm \delta T})$ may be indicative of the interference of waves with the plume dynamics, which may undermine the criterion presently used to define the depth of a plume. We will analyse in more details these effects in two follow-up studies devoted to internal waves (Le Saux et al., in prep) and to  the analysis of mixing in the overshooting layer based on tracer particles (Guillet et al., in prep).

\begin{figure}[h!]
\vspace{0cm}
\includegraphics[height=8cm,width=8cm]{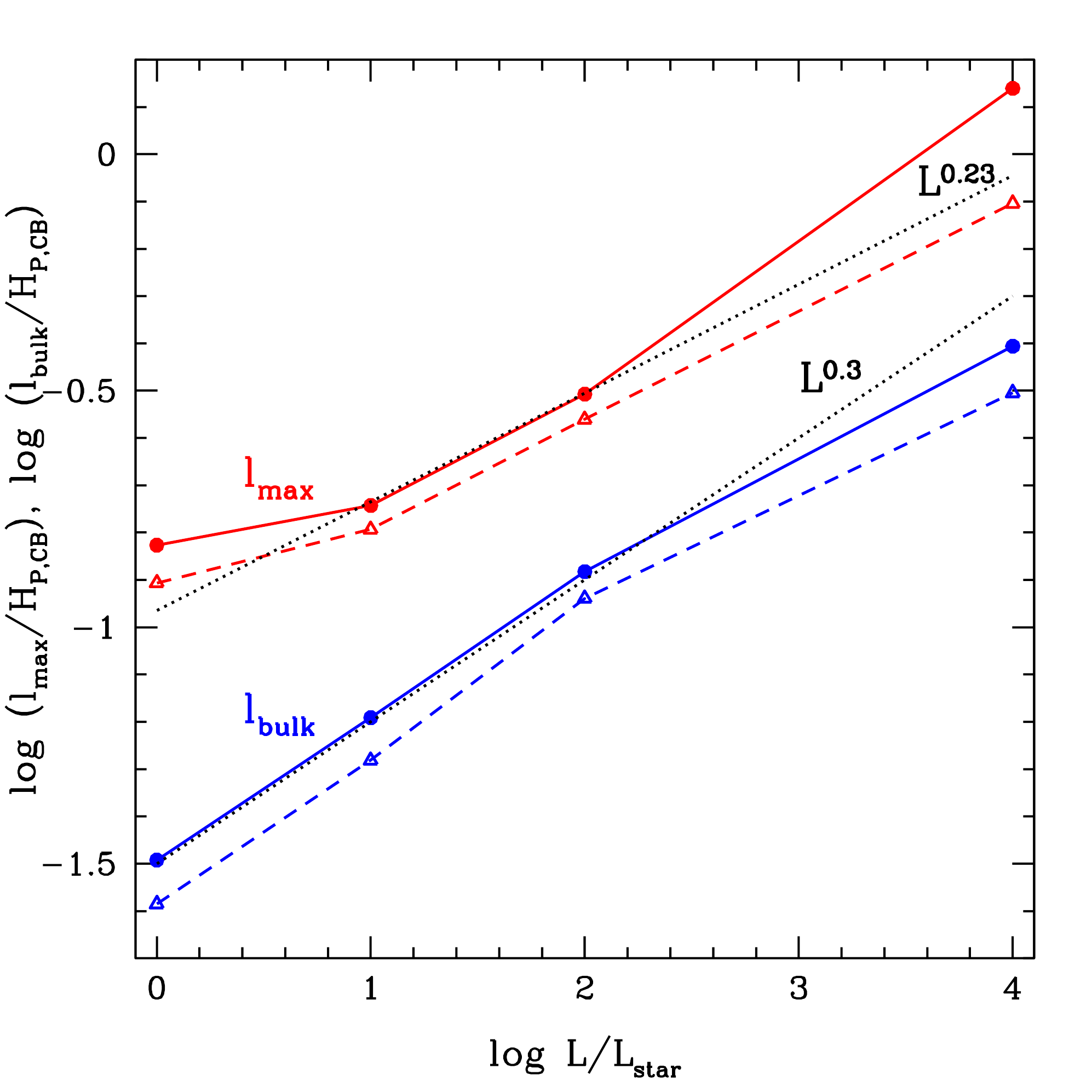}
\vspace{-0cm}
   \caption{Maximal penetration length $l_{\rm max}$ (red) and length of the layer where the bulk of the plumes penetrate $l_{\rm bulk}$ (blue curves), in units of the pressure scale height at the convective boundary, as a function of the luminosity of the model. The solid curves and circles are the values derived from the use of $\mathbf f_{\rm k}$ and the dashed curves and triangles corresponds to the values derived from $\mathbf f_{\rm \delta T}$. The dotted curves show power laws $L^{0.23}$ and L$^{0.3}$ as guide for the eyes.}
 \label{lov_fig}
\end{figure}

Our data are not extensive enough to robustly determine a power law scaling of the overshooting depth with the luminosity of the model,
as done in e.g. \ct[][]{hotta17} and \ct{ kapyla19}. It is however still interesting to examine the scaling suggested by our results. This is illustrated in Fig. \ref{lov_fig} which shows the results for both $l_{\rm bulk}$ and $l_{\rm max}$. The results for models boost1d4 are also shown for illustration in Fig. \ref{lov_fig}, but they
 should not be used to infer a scaling since the penetration layer progresses deeper with time.
Scaling of the overshooting depth $l_{\rm ov}$ reported in the literature from numerical simulations can vary from $l_{\rm ov} \propto L^{0.08}$ to 
 $l_{\rm ov} \propto L^{0.31}$ and strongly depends on the heat conductivity profile at the convective boundary, as shown by \ct{kapyla19}. The scaling also depends on the method used to define $l_{\rm max}$, as illustrated in Fig. \ref{lov_fig}. The results for $l_{\rm max}$ do not show a simple scaling law with $L$. The results for  $l_{\rm bulk}$ show an approximate scaling $\propto L^{0.3}$. This suggests that the dynamics of the bulk of the plumes is primarily  linked to the convective velocity  $v_\mathsf{rms} \propto L^{1/3}$, whereas the determination of $l_{\rm max}$ could be  impacted by other factors,
 which will be the focus of a follow-up study based on tracer particles. In the next section we analyse in more details the statistics of key quantities in the overshooting layer to gain more insight about the properties of this region. 
 The results for overshooting lengths, listed in Table \ref{tabdepth}, correspond to an initial model that has differences from a solar model, as explained in \S2.1.  These  lengths therefore should not be used more generally to characterise the overshooting in other stars.
 
\begin{figure*}[h]
\hspace{-0.5cm}
\includegraphics[height=7cm,width=6.5cm]{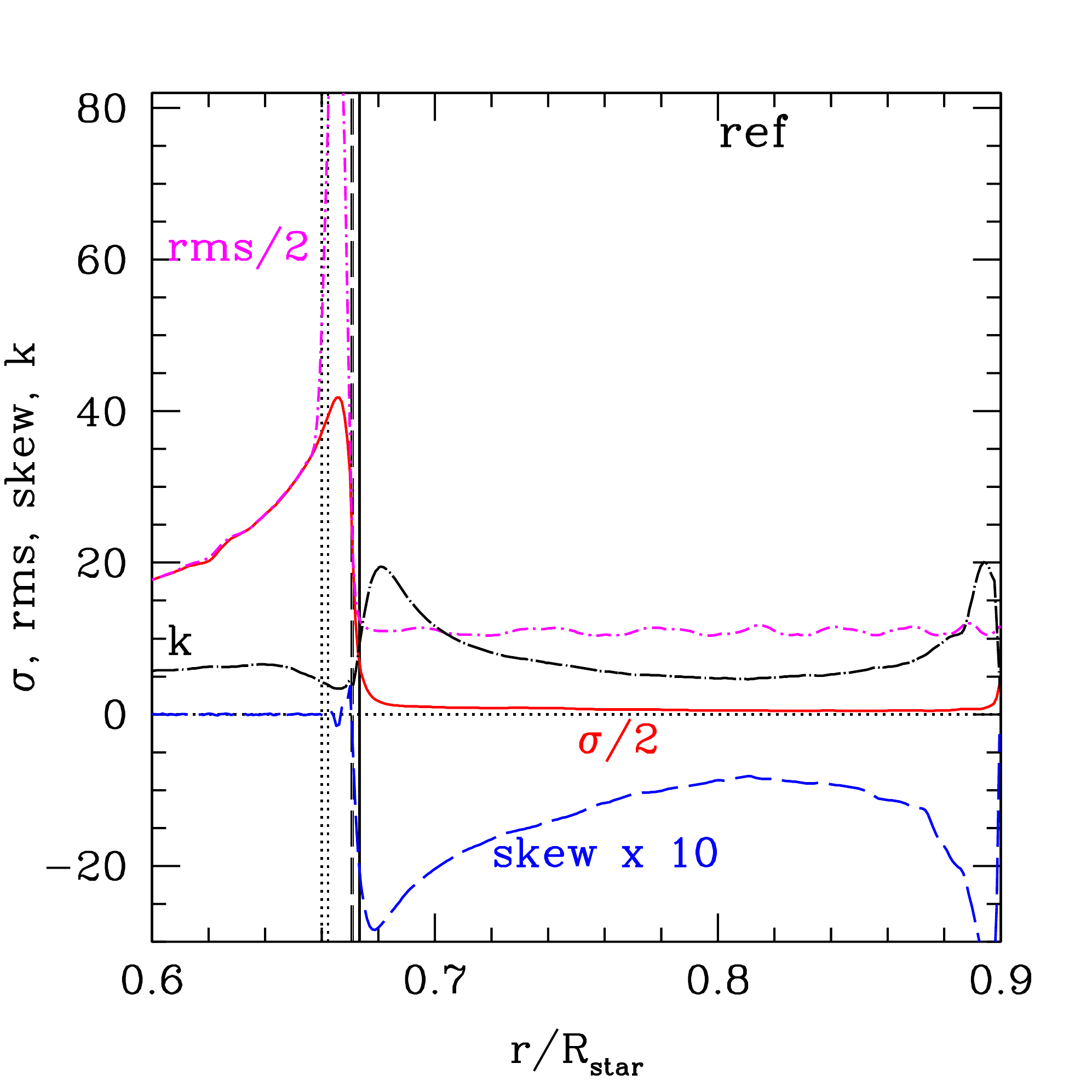}
\hspace{-0.5cm}
\includegraphics[height=7cm,width=6.5cm]{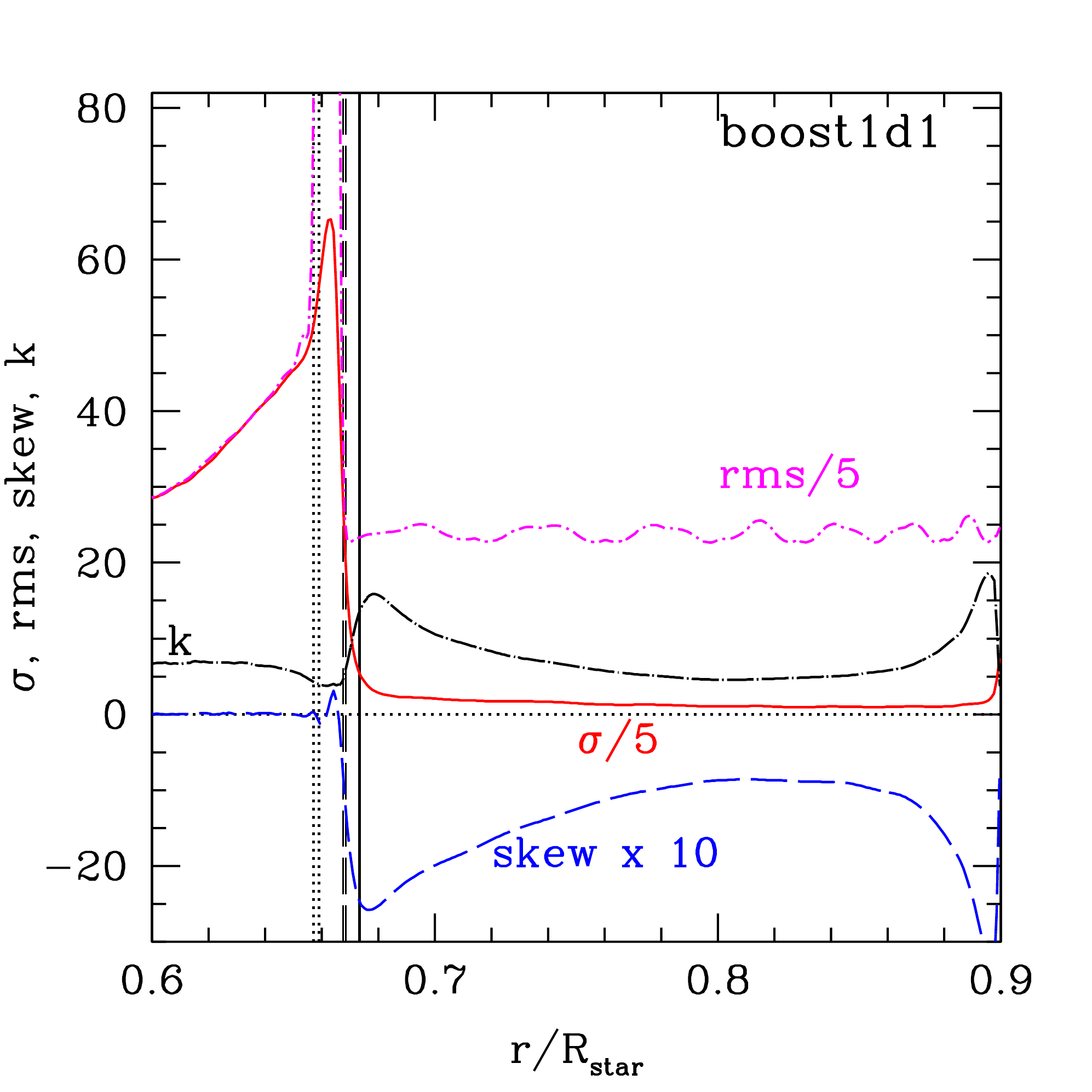}
\includegraphics[height=7cm,width=6.5cm]{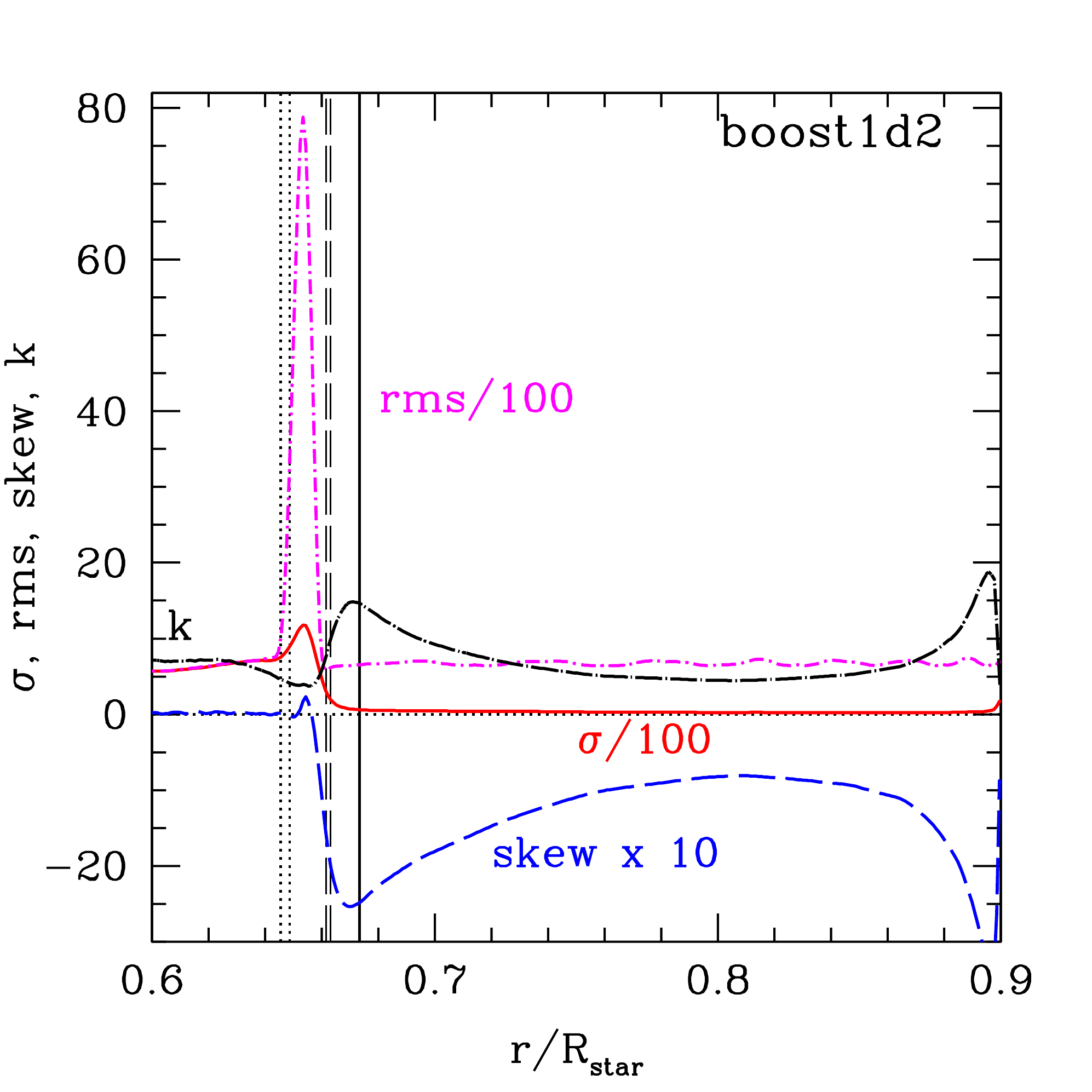}
\hspace{-0.35cm}
   \caption{Time average of standard deviation (solid red), rms (dash-dot magenta), skewness (long dash blue) and kurtosis (long dash-dot black) of the temperature fluctuations for the models indicated
 in the panels. The convective boundary corresponding to the Schwarzschild boundary from the 1D initial model is indicated by the vertical solid line. The two vertical dotted lines in each panel indicate the extension of the penetration layer $l_{\rm max}$ based on $\mathbf f_{\rm k}$ and $\mathbf f_{\rm \delta T}$, respectively. Similarly, the two vertical dashed lines correspond to $l_{\rm bulk}$.}
 \label{avsdt_fig}
\end{figure*}

\section{Results: Statistical characteristics of the overshooting layer}
\label{fluctuations}
A more in-depth analysis of the temperature fluctuations and associated second, third and fourth-order moments can provide additional information about the flow structure in the convection and penetration regions. It also provides additional elements to understand our results and their interpretation. Because of the unsteady evolution of model boost1d4, this model is excluded from a high-order moment analysis. 
For the three remaining models,  we calculate the standard deviation, rms, skewness and kurtosis of the temperature fluctuation, defined at time $t$ and at each radius $r$ by:
\begin{eqnarray}
\mathsf{\sigma}_{\rm \dt}(r,t) &=&   \sqrt{\big \langle (\delta T -   \langle \delta T \rangle_\theta)^2 \big \rangle_{\theta}} , \label{sigma} \\ 
\mathsf{rms}_{\rm \dt}(r,t) &=&    \sqrt{\langle \delta T^2 \rangle_{\theta}} , \label{rms} \\ 
\mathsf{skew}_{\rm \dt}(r,t) &=&    { \big \langle(\delta T -  \langle \delta T \rangle_\theta)^3 \big \rangle_{\theta}  \over \sigma^3_{\rm \dt}}, \label{skew}  \\ 
\mathsf{kurt}_{\rm \dt}(r,t) &=&   { \big \langle(\delta T -  \langle \delta T \rangle_\theta)^4 \big \rangle_{\theta}  \over \sigma^4_{\rm \dt}}. \label{kurt} 
\end{eqnarray}
The temperature fluctuation $\delta T$ is defined by Eq.(\ref{eqdelta}). Figure {\ref{avsdt_fig} shows  the time average of these quantities.
The skewness reflects the asymmetry of the temperature fluctuation distribution around the mean value. 
The kurtosis provides a measure of the  significance of extreme temperature fluctuations in the temperature distribution.
In the convective zone, the skewness is negative due to the descending cold plumes.
At the bottom of the convective zone the kurtosis increases in amplitude, indicating stronger temperature deviations. 
This behaviour is also due to the cold downflows which are immersed in increasingly hotter background as they descend towards the convective boundary. This is confirmed by the negative bump of the skewness where the kurtosis has its maximum. 
We have indicated in Fig. {\ref{avsdt_fig} the position of $l_{\rm bulk}$ and $l_{\rm max}$. The position of the negative peak of the enthalpy flux coincides with the level $l_{\rm bulk}$ where the plumes penetrate frequently. This confirms the interpretation of this negative peak (see \S \ref{dynamics}). It is interesting to note that the general behaviour of these high-order moments is very similar across the range of luminosity enhancement
factors, both within the convective zone and at the convective
boundary.

For the three models, the strongest departure between $\mathsf{\sigma}_{\rm \dt}$ and $\mathsf{rms}_{\rm \dt} $ lies between the location where the bulk of the plumes penetrates and the layer reached by the extreme plumes. The difference between  $\mathsf{\sigma}_{\rm \dt}$ and $\mathsf{rms}_{\rm \dt}$ is due to the time evolution of the mean temperature profile. Indeed if $ \langle T \rangle_\theta = \big \langle {\langle T \rangle_\theta} \rangle_t$, then the standard deviation and the rms are equivalent. This strong departure thus indicates the modification of the temperature background  due to the slight heating mentioned in \ref{heating}. The depth reached by the most vigorous plumes $l_{\rm max}$ coincides with the position where  $\mathsf{\sigma}_{\rm \dt}$ and  $\mathsf{rms}_{\rm \dt}$ get closer again, i.e. where the background is not significantly evolving.  A closer look at the penetration layers shows that the position of $l_{\rm max}$ is located in the region where the subadiabaticity $|\nabla - \nabla_{\rm ad}|$ increases compared to the initial profile, implying more efficient braking of descending downflows. This is illustrated in Fig. \ref{cor_fig} which shows the profile of $\big \langle { \langle (\nabla - \nabla_{\rm ad}) \rangle_{\theta}} \big \rangle_t - (\nabla - \nabla_{\rm ad})_{\rm init}$ for the three models.  This is consistent with the picture described in \S \ref{heating} and highlights the importance of the thermal profile modified by the penetrating plumes and affecting in return their dynamics and the effective overshooting depth. 

The connections found between the position of $l_{\rm max}$ and the behaviour of  $\mathsf{\sigma}_{\rm \dt}$ and  $\mathsf{rms}_{\rm \dt}$ and between $l_{\rm max}$ and the subadiabaticity are of particular interest and will be the focus of future studies as these quantities may provide other criteria to determine the depth of the overshooting layer. The statistical approach performed in \S \ref{statistics} requires simulation times covering several hundreds of convective turnover timescales ($\simgt$ 400 $\times$ $\tau_{\rm conv}$) to reach statistical convergence. Criteria based on the behaviour of $ \mathsf{\sigma}_{\rm \dt}$ and  $\mathsf{rms}_{\rm \dt}$ and/or on  the profile of ($\nabla - \nabla_{\rm ad}$) would not require such long simulation times and would provide an advantageous alternative  approach, in terms of computation time.

\begin{figure}[h!]
\vspace{-0.cm}
\includegraphics[height=11cm,width=9cm]{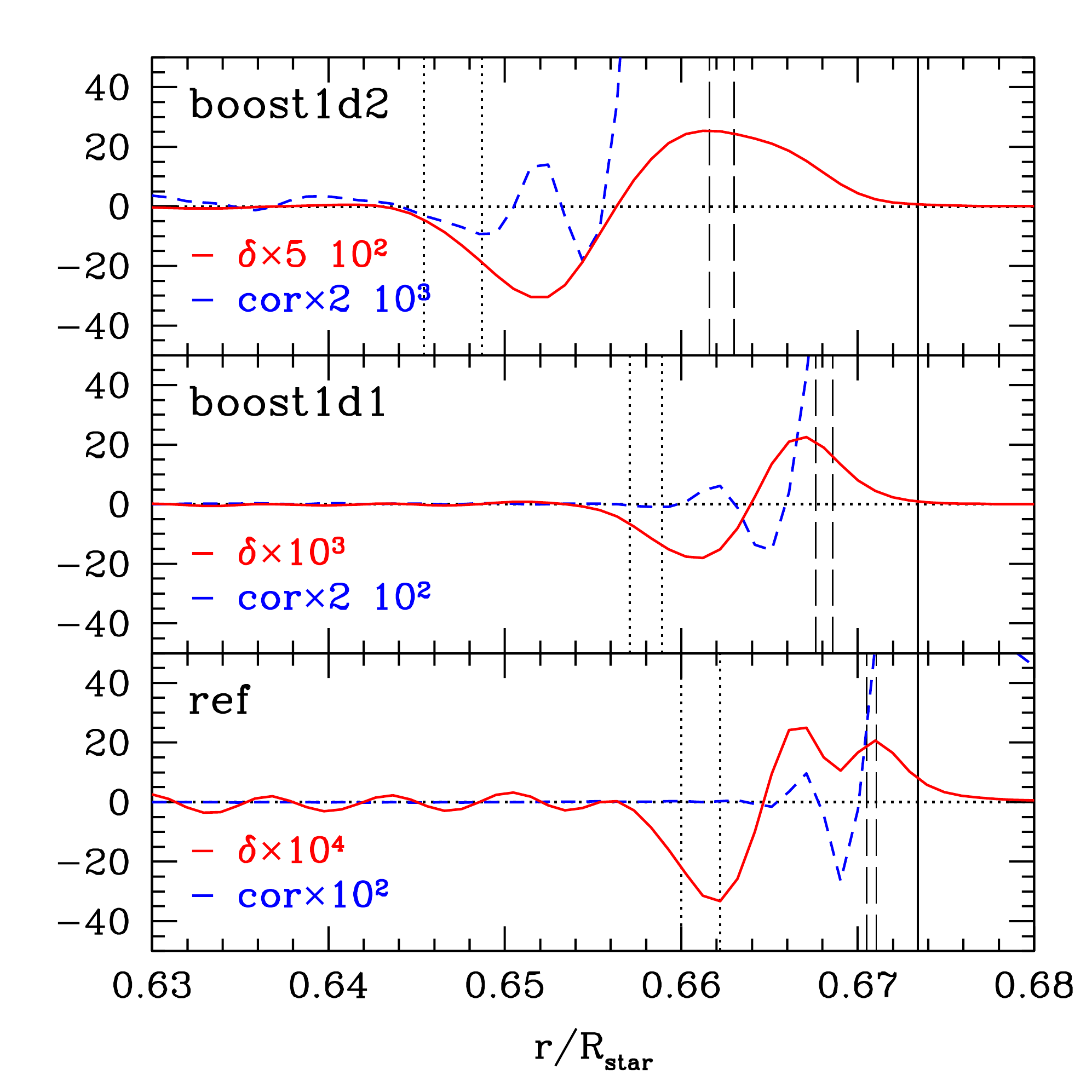}
\vspace{-0cm}
  \caption{Solid lines (red): Departure of the  subadiabaticity (time and horizontal average) from its initial value $\delta = \big \langle { \langle (\nabla - \nabla_{\rm ad}) \rangle_\theta} \big \rangle_t - (\nabla - \nabla_{\rm ad})_{\rm init}$ as a function of radius in the overshooting region for the three models, as indicated in the panels.  Dashed lines (blue): radial profile of $\langle {\mathsf{cor}} \rangle_t$, the time average of the radial velocity autocorrelation defined by Eq. (\ref{eq_cor}). Since these quantities are very small, they are multiplied by different constants, as indicated in each panel, in order to see the details.
  The vertical lines are the same as in Fig. \ref{avsdt_fig}.}
 \label{cor_fig}
\end{figure}

 \ct{korre19} and \ct{cai20} suggested to use the radial velocity correlation with the velocity at the convective boundary as an alternative measure of the extent of the overshooting distance. They argue that this quantity  should favour the strongest downflows and  capture  extreme events. \ct{cai20} suggests that the second-zero of this velocity autocorrelation is a good indicator for the measurement of upward overshooting.  We examine this suggestion and, as \ct{cai20}, we define the velocity autocorrelation at a given time and at each radius $r$ by:
\begin{equation}
\mathsf{cor}(r,t) =  {\big \langle \velvec_r (r,\theta,t) \velvec_{r} (r_{\rm conv},\theta,t) \big \rangle_\theta  \over \vel_{r,\mathsf{rms}}(r,t) \vel_{r,\mathsf{rms}} (r_{\rm conv},t) }, \label{eq_cor}
\end{equation}
with $\vel_{r,\mathsf{rms}}$ the rms of the radial velocity following the definition given in Eq.(\ref{eq_rms}) and $r_{\rm conv}$ the radius at the convective boundary. We show the radial profile of the time average $\langle {\mathsf{cor}} \rangle_t$  in Fig. \ref{cor_fig} and compare it to the position of $l_{\rm max}$ and the profile of the subadiabaticity. We find that the velocity autocorrelation has several zeros between $l_{\rm bulk}$ and $l_{\rm max}$. The second zero lies between these two lengths, but its position shows no systematic link with the position of $l_{\rm max}$ and with the behaviour of the subadiabaticity. A more thorough analysis of the most relevant criterion to characterise the mixing in the overshooting layer and its length is beyond the scope of this study and will be the focus of a follow-up study. 

\section{Discussion and conclusion}
\label{conclusion}

The experiments with artificially enhanced luminosity and thermal diffusivity performed in this study have several implications. Our results suggest a link between the properties of penetrating plumes and the thermal structure of the overshooting layer; this confirms the findings of
earlier works  \cp[e.g.][]{hurlburt86, muthsam95, rempel04, kapyla19}.
They stress the importance of the impact of the penetrative downflows on the local thermal background, as a result of compression and shear which induce local heating and thermal mixing in the overshooting layer. This effect is not pronounced in the ref model, but is present. The artificial increase in the energy flux intensifies the effect by increasing the velocities in the convective zone and at the convective boundary. This artefact reveals the subtle connection between the local heating of the thermal background and the plume dynamics as the heating alters the subadiabaticity in the penetration region. This local heating  increases also the efficiency of the heat transport by radiation, which may counterbalance further heating and help establishing a steady state. We suggest that the modification of the thermal background by the penetrative plumes and its impact on the radiative flux and on the plume dynamics play a role on the efficiency of the mixing driven by the  plumes and eventually on the width of the overshooting layer.  

This finding suggests that an artificial modification of the radiative diffusivity in the overshooting layer, rather than only accelerating the thermal relaxation, may also alter this balance and most likely affect the efficiency of the overshooting mixing and thus the effective width of the overshooting layer. In the most boosted model presently studied, namely boost1d4, this balance seems difficult to establish because of the strong plumes driving efficient heating and progression of the penetration layer deeper inwards. In the context of the study of overshooting, the argument that a simulation with enhanced energy flux and radiative diffusivity compared to a reference model is an accelerated version of this model, or in other words that the boost1d4 simulation at a given time provides a good representation of the ref simulation at much later times, is difficult to justify in light of present results. 

Admittedly, stellar hydrodynamic simulations based on realistic energy flux and radiative diffusivity still remain very far away from realistic stellar conditions. 
For models with no or moderate enhancement of the luminosity, the numerical diffusion due to truncation errors likely dominates the physical radiative diffusion. Consequently, the effective P\'eclet number, which measures the relative importance of advection versus diffusion, is set by the numerical diffusion and thus underestimated. For large enhancement of the luminosity and of the radiative diffusivity, as in the boost1d4 model, the physical radiative diffusion can start to dominate over the numerical diffusion. In these models, the heat diffusion is hence better described. But unfortunately,  these models drift away strongly from their initial state and their
relaxed state will not describe the initial one.
Artificial enhancement of the energy flux or modification of the thermal diffusivity profile therefore pushes the simulated conditions away from the original target star. Based on present results, we do not advise using luminosity enhancement factors $\simgt 10^4$ as they induce a significant drift from the initial stellar structure. Factors even larger ($> 10^6$) could produce unrealistic supersonic velocities close to the stellar surface.

The modification of the local thermal background in the overshooting region has  been reported in several works \cp[e.g.][]{rogers06, viallet13, korre19, higl21}, but has not been the subject of deeper scrutiny to understand its origin.  This lack of deepening is most likely due to the fact that many of these simulations, including our work, have not achieved thermal relaxation, raising a concern about the physical relevance of this feature. Paradoxically, many numerical simulations acknowledge the existence of a bump of the radiative flux in the penetration region \cp[e.g.][]{brun11, brun17, hotta17, kapyla19, cai20}, needed to counterbalance the negative enthalpy flux. But no details are given regarding what causes this bump, except that it is there to ensure thermal equilibrium. There must be physical processes which yield such a peak in the radiative flux. Our simulations also feature this bump (see Fig. 4). Several thermal relaxation timescales are required in order for the radiative flux to compensate {\it exactly} for the negative enthalpy flux. Our reference and moderately boosted models are not able to reach this state, whereas as discussed above, the most boosted simulation is evolving towards an extreme state. Because our experiments link the radiative flux bump to the modification of the local thermal background due to the penetrating downflows, they could help understanding the underlying physical processes responsible for this bump.


The sensitivity of this feature to some of the input physics included in the models or to the dimensionality of the numerical simulations should be examined.
In order to test the robustness of the local heating process against the equation of state, we constructed an artificial 1D stellar model close to the stage of evolution of the ref model but with an ideal gas equation of state with constant adiabatic index $\gamma = 5/3$. Such a simple equation of state is implemented in the MUSIC code and is straightforward to implement in a 1D stellar evolution code. This provides an unrealistic stellar model because $\gamma$ varies in the stellar interior due to recombination or ionisation processes in the convective envelope and in the overshooting region.  Indeed, the radiative-convective transition at the bottom of the convective envelope corresponds to the last bump in opacity due to partial ionisation of metals \cp{rogers92}. However, such a model provides an interesting case to test the sensitivity of the heating process to the equation of state and, for the most boosted model, the fast inward progression of the penetration layer. We ran two cases with the MUSIC code starting from this artificial stellar model with luminosity enhanced by a factor 10 and 10$^4$, respectively,
and find the same modification of the thermal profile and same heating in the overshooting region, and for the most boosted model, the same inward progression of the penetration layer with time as found for models with a realistic stellar equation of state. 

We performed an additional test to check the robustness of the results against the treatment of gravity. 
Given the non-negligible modification of the background profile found in the boosted models, we have verified for the boost1d1 and boost1d4 models that updating the calculation of the spherically symmetric gravitational acceleration $\vec g$ during the simulation has no impact on the dynamics of the plumes and on the heating process. Since there is no need to re-calculate $\vec g$ at each numerical timestep, it was re-calculated every time interval $\Delta t$, with $\Delta t$ fixed during a whole test simulation. The typical dynamical timescale of the stellar model $\tau_{\rm dyn} \sim 1/\sqrt{(\rho_{\rm mean} G)}$, with $\rho_{\rm mean}$ the mean density and $G$ the gravitational constant, is of the order of 10$^3$s. We have thus performed a number of test simulations with $\Delta t$ varied between $10^2$ and $10^5$ seconds and found no impact on our main results.
 
One may also question the impact of the resolution and of the dimensionality on the local modification of the thermal background. We performed two test simulations for boost1d1 and boost1d4 with double resolution (1024x1024) over a few convective turnover timescales and find the same local heating and modification of the thermal background over similar timescales compared to the 512x512 counterpart. Even though such resolution tests will not bring the simulations close to the true parameter regime, they remain useful as they provide a test for the sensitivity of the results to the numerical diffusion.

We are currently performing three-dimensional simulations of a solar-like model but they are not yet exploitable to analyse the time evolution of the thermal background in the overshooting layer. In the meantime, we can only speculate on the impact of dimensionality on these results. \ct{brummell02} find different structures of the penetrative convection in 3D compared to 2D simulations. They still find significant overshoot of the convective motions in highly turbulent 3D simulations, but these motions do not establish an adiabatic penetration region as previously found in 2D simulations \cp[e.g.][]{hurlburt94}. Even if the structure and the geometry of the penetrating downflows are modified in 3D, these motions should still produce compression and shear in the overshooting region. We should thus still expect them to induce local heating and thermal mixing, and consequently a modification of the thermal background.
We note that the alteration of the thermal background has been reported in other numerical simulations based on very different numerical approaches, namely a 2D anelastic method in \ct{rogers06} and a 3D approach within the  Boussinesq approximation in \ct{korre19}. The resolution and dimensionality may however impact the establishment of the balance between the local heating, the increase in the radiative flux and the inward progression of the penetration layer. We suspect that this balance is subtle in numerical simulations and could be impacted by various factors and uncertainties. We are not able to be more specific based on the present study and this is the subject of future experiments we are performing for convective envelopes and convective cores.

Simulations with an artificial increase in the energy flux may be of limited relevance for the quantitative determination of an overshooting distance. Because of additional artefacts introduced for large enhancement factors, scaling laws based on such simulations need to be applied with caution.
But such simulations
are still valuable experiments to perform for moderate enhancement factors. 
They may be used for the general analysis of convective flow structure based on high-order statistics of  temperature fluctuations, given the similarities found across a range of modest luminosity enhancement factors.
These experiments may also be used for the predictions of internal wave spectrum and properties, modulo an appropriate rescaling. This has never been proven or disproven and is the purpose of a follow-up study (Le Saux et al., in prep). They also provide excellent laboratories to compare a method based on fluxes in an Eulerian approach and a method based on Lagrangian tracer particles to define an overshooting depth characterising material mixing.

Lastly, as suggested by e.g. \ct[][]{rempel04} and \ct[][]{zhang12}, the modification of the thermal background at the convective boundary due to convective penetration  could improve the sound speed discrepancy between helioseismology data and  solar models. Our results suggest a complex behaviour of the temperature gradient at the convective boundary that deserves further attention, not only in the context of helioseismology but also for the general understanding of convective penetration.   

\section{Acknowledgements}
This work is supported by the ERC grant No. 787361-COBOM and the consolidated STFC grant ST/R000395/1. IB thanks the Max Planck Institut f{\"u}r Astrophysics (Garching) for warm hospitality during completion of part of this work. The authors would like to acknowledge the use of the University of Exeter High-Performance Computing (HPC) facility ISCA and of the DiRAC Data Intensive service at Leicester, operated by the University of Leicester IT Services, which forms part of the STFC DiRAC HPC Facility. The equipment was funded by BEIS capital funding via STFC capital grants ST/K000373/1 and ST/R002363/1 and STFC DiRAC Operations grant ST/R001014/1. DiRAC is part of the National e-Infrastructure.

\bibliographystyle{aa.bst}
\bibliography{references}

\end{document}